\documentclass[aps,prd,twocolumn,preprintnumbers,superscriptaddress,nofootinbib,floatfix]{revtex4}

\usepackage{multirow, graphicx,amssymb,url,mathrsfs,amsmath}
\usepackage{eucal,wrapfig,boxedminipage,setspace,subfigure}
\usepackage{amsxtra,amstext,latexsym,dsfont}
\usepackage[colorlinks=true,linkcolor=blue,citecolor=blue]{hyperref}
\usepackage{comment}

\newcommand{\cala}{\mbox{${\cal A}$}}

 \newcommand{\call}{\mbox{${\cal L}$}}

\def\IR{{\hbox{{\rm I}\kern-.2em\hbox{\rm R}}}}
\def\IB{{\hbox{{\rm I}\kern-.2em\hbox{\rm B}}}}
\def\IN{{\hbox{{\rm I}\kern-.2em\hbox{\rm N}}}}
\def\IC{\,\,{\hbox{{\rm I}\kern-.59em\hbox{\bf C}}}}
\def\IZ{{\hbox{{\rm Z}\kern-.4em\hbox{\rm Z}}}}
\def\IP{{\hbox{{\rm I}\kern-.2em\hbox{\rm P}}}}
\def\IH{{\hbox{{\rm I}\kern-.4em\hbox{\rm H}}}}
\def\ID{{\hbox{{\rm I}\kern-.2em\hbox{\rm D}}}}

\def\del{\partial}

\def\det{{\rm det}}

\def\nn{\nonumber}

\newcommand{\beq}{\begin{equation}}
\newcommand{\eeq}{\end{equation}}
\newcommand{\bea}{\begin{eqnarray}}
\newcommand{\eea}{\end{eqnarray}}

\newcommand{\dd}{\mathrm{d}}
\usepackage{ulem}

\newcommand{\nothing}[1]{}

\begin{document}

\voffset 1cm

\newcommand\sect[1]{\emph{#1}---}

\title{Properties of Stable Massive Quark Stars in Holography}

\author{Kazem Bitaghsir Fadafan}
\affiliation{ Faculty of Physics, Shahrood University of Technology,
P.O.Box 3619995161 Shahrood, Iran}

\author{Jes\'us Cruz Rojas}
\affiliation{Departamento de F\'isica, Facultad de Ciencias, Universidad
Nacional Aut\'onoma de M\'exico, Apartado Postal 70-543, CDMX 04510, M\'exico}

\author{Jonas Mager}
\affiliation{Institut f\"ur Theoretische Physik, Technische Universit\"at Wien,
        Wiedner Hauptstrasse 8-10, A-1040 Vienna, Austria}

\begin{abstract}
We study a holographic D3/D7 system, whose dilaton profile has been phenomenologically adjusted in the infrared. The model is used to describe a deconfined yet massive quark phase of QCD at finite density, concluding that the equation of state of such a phase can be stiff enough to support exotic dense stars as massive as 2 solar masses. Nucleons are modeled phenomenologically using the Hebeler-et.al EFT baryon phases. For the stiff phenomenological baryon phases the transition to the quark phase is weakly first order allowing for stable quark cores. We also find that holographic baryons, modeled as wrapped D5-branes, provide unrealistic pressures (in the homogeneous approximation) and have to be discarded. We compute the mass vs. radius relation and tidal deformability for these hybrid stars. Contrary to a large number of other holographic models, this holographic model indicates that quark matter could be present at the core of heavy compact stars and may be used to explore the phenomenology of such objects.

\end{abstract}

\maketitle

\newpage

\section{Introduction}\label{SectionI}

Quark stars, a class of compact stars, are believed to form under extreme conditions where the core density is so high that nucleons break down into their constituent quarks, resulting in quark matter \cite{Ivanenko:1965dg,Itoh:1970uw,Alcock:1986hz,Witten:1984rs}. A hadron-quark phase transition in the dense cores of neutron stars is considered a plausible mechanism for the formation of quark matter. These stars could emerge from the remnants of supernovae, transitioning from neutron stars if the conditions are sufficiently extreme. 
Several studies have explored whether quark stars can be formed in nature; in \cite{Annala:2023cwx} the authors combined information from astrophysical observations and calculations using Effective Field Theory (EFT) and perturbative QCD to argue that in the cores of maximally massive stars, the Equation of State (EoS) is consistent with quark matter. There are also studies using perturbative QCD, which have found that dense quark stars are indeed plausible \cite{Fraga:2001id, Kurkela:2010yk, Kopp:2025ggp}.

Recent observations of compact objects through gravitational wave (GW) signals \cite{LIGOScientific:2017vwq} (and their electromagnetic counterparts), mass measurements of
the heaviest pulsars \cite{Demorest:2010bx, Antoniadis:2013pzd, Fonseca:2016tux, NANOGrav:2019jur, Fonseca:2021wxt} and mass-radius measurements using X-ray observations by the NICER
and other collaborations \cite{Nattila:2017wtj, Miller:2019cac, Miller:2021qha}, have provided a large amount of data for compact stars. This data places constraints on the dense matter EoS, hinting at the possibility that cores of very massive neutron stars contain a deconfined quark matter phase rather than just a baryonic phase \cite{Annala:2023cwx}. In this case, the highly compressed nuclear matter phase is expected to undergo a phase transition to deconfined quark matter, releasing its constituent quarks and gluons. Consequently, the EoS of quark matter may be crucial in determining the structure of the star beyond nuclear saturation density. Anticipated advancements in multi-messenger astrophysics offer promising pathways to explore this question \cite{Evans:2021gyd, ET:2025xjr}.

In the astrophysics community one standard way to approximate the EoS is via the MIT bag model \cite{Chodos:1974je}. Within the framework of general relativity, one may subsequently explore its impact on the key macroscopic characteristics of compact stars, such as mass and radius. 
Within the context of modified gravity the static stability criterion, the adiabatic index, and the sound velocity were studied in \cite{Yagi:2016bkt}. However, there are many limitations in such approaches. In the strongly coupled regime  there is currently no first principles calculation that can provide the EoS. Despite theoretical hints, quark stars remain hypothetical. Observational challenges, such as their potential resemblance to purely hadronic neutron stars and the uncertainties surrounding the EoS for quark matter, need to be faced to make progress.

First-principles methods, such as lattice QCD, are not well-suited for addressing the regime where quark matter is expected to appear due to the fermion sign problem at large densities. On the other hand, traditional phenomenological approaches such as nucleon EFTs often fail to describe those regimes due to the extreme conditions present in neutron stars where QCD becomes strongly coupled. 
These problems motivate the exploration of alternative non-perturbative approaches. Holographic models have been used to tackle other challenging problems in QCD, very recently for example in the context of hadronic light-by-light scattering \cite{Leutgeb:2022lqw,Leutgeb:2024rfs,Cappiello:2025fyf,Mager:2025pvz}. In this work we analyze the predictions of particular models of holographic QCD (hQCD) which originate from phenomenological extensions of D3/D7-brane configurations in AdS/CFT.

hQCD models are inspired by the  gauge-gravity duality which relates large $N_c$ strongly coupled QFTs to weakly coupled gravitational theories in higher dimensions. The challenging problem of computing the EoS for a strongly coupled theory like QCD can therefore be approached by working in the dual higher dimensional theory of gravity.
Holographic models thus allow one to tackle the strongly coupled regime of matter and predict observable properties of compact stars like the mass-radius relation and cooling behaviors.

Within holography, one can differentiate between top-down and bottom-up models. Top-down models originate from specific, often ten-dimensional, string theory brane constructions where the dual gauge theory is well understood \cite{Witten:1998zw, Karch:2002sh, Sakai:2004cn, Sakai:2005yt}. Bottom-up models usually follow the general lessons learned from top down approaches, but there is considerably more freedom in choosing e.g., field content, actions, etc. While the global symmetries and operator content of QCD provide guidance for model building, the bottom-up models lack a precise correspondence to a known field theory. As a result, they contain adjustable parameters that must be tuned to reproduce QCD phenomenology. Compared to top-down approaches the added flexibility is very useful and often necessary to describe important aspects of QCD realistically. In this paper we employ a bottom-up model which is heavily inspired by rigorous D3/D7-brane constructions \cite{Karch:2002sh} in IIB string theory.

Among the plethora of bottom-up holographic models, which include the hard- and soft-wall\footnote{Soft-wall models in their most basic form, as well as in a large class of deformations have recently been shown to lead to divergent 4-point functions of QCD flavor currents \cite{Leutgeb:2025jmv}. } models \cite{Erlich:2005qh, DaRold:2005mxj, Karch:2006pv}, and models based
on Einstein-Maxwell actions as well as more sophisticated models such as the V-QCD model \cite{Jarvinen:2011qe}, bottom-up models based on the D3/D7 configuration in IIB string theory are especially suitable to describe compact stars \cite{BitaghsirFadafan:2019ofb, BitaghsirFadafan:2020otb}. They have enough room for adding ingredients that make the dual theory more realistic, but they are not overly complicated numerically. In this work, we use this kind of model, with a specific non-trivial dilaton profile.

In the holographic D3/D7 setup, the dilaton — a scalar field that determines the coupling in string theory — is chosen to have a phenomenological background profile (while neglecting backreactions to the metric in a first approximation) modeling the running coupling of the dual gauge theory. This modification is also useful in studies of chiral symmetry breaking \cite{Evans:2010iy}. Such profiles are designed to interpolate between ultraviolet (UV) and infrared (IR) behaviors, offering a more realistic depiction of the gauge coupling’s energy dependence.

For instance, a dilaton that transitions smoothly from linear in the UV to exponential in the IR can effectively model phenomena such as chiral phase transitions. By tuning its parameters, a wide range of strongly coupled dynamics can be explored. This method draws inspiration from top-down constructions: a well-known type IIB supergravity solution with features of confinement was introduced in \cite{Gubser:1999pk}, while non-supersymmetric backgrounds with deformed AdS geometry and a running dilaton were developed in \cite{Constable:1999ch} and applied to chiral symmetry breaking in \cite{Babington:2003vm}. These efforts highlight how tailored dilaton profiles help capture essential aspects of gauge theories, such as confinement and chiral symmetry breaking, deepening our understanding of strongly coupled systems.
In \cite{BitaghsirFadafan:2019ofb}, a restrictive set of dilaton profiles was considered, which additionally had a divergence in the deep IR. The scalar quasinormal modes of this system at finite temperature have been studied in \cite{Atashi:2024jbk}. The behavior of the critical point of QCD-like theories and magnetically induced composite inflation within D3/D7 models have been investigated in \cite{BitaghsirFadafan:2024icz,Ahmadvand:2021wtt}. The present work aims to extend analysis of the equations of state to a wider range of profiles for D3/D7 models which are regular in the IR.

Applications of holography to the EoS at nonzero baryon chemical potential and the properties of neutron stars have a rich history \cite{Bergman:2007wp,Rozali:2007rx,Kim:2007vd,Kaplunovsky:2012gb,Li:2015uea}. Neutron stars and the question whether or not stable quark cores can exist inside of them were investigated in holography within the Sakai-Sugimoto model \cite{Elliot-Ripley:2016uwb,Kovensky:2020xif,Preis:2016fsp,BitaghsirFadafan:2018uzs,Kovensky:2021kzl,Bartolini:2023wis}, the VQCD-model \cite{Jokela:2018ers,Chesler:2019osn,Ecker:2019xrw,Jokela:2020piw,Jarvinen:2021jbd,Jarvinen:2023jbr,Bartolini:2025sag} as well as hard- and soft-wall models \cite{Bartolini:2022rkl,Zhang:2022uin}. In these holographic models quark stars are more or less strongly disfavored.

The study of neutron stars in D3/D7 type models was initiated in \cite{Hoyos:2016zke} in which a constant dilaton, corresponding to the original $\mathcal{N}=4$ SYM setup with probe D7-branes, was utilized (see also \cite{Aleixo:2023lue}). In this model no spontaneous chiral symmetry breaking exists, but a quark mass parameter allows for explicit breaking. The EoS \cite{Karch:2007br} describes a massive deconfined quark phase. It was evident in this first attempt that deconfined conformal matter from holographic models based on the pure D3/D7 case do not yield an EoS which is stiff enough to obtain very massive compact stars.

In \cite{BitaghsirFadafan:2019ofb} the D3/D7 system was modified to incorporate a running anomalous dimension for the quark condensate. This adjustment introduces a dynamic mechanism for chiral symmetry breaking while maintaining a deconfined massive phase at intermediate densities. Interestingly, it was shown that this produces significantly stiffer equations of state and exhibits non-monotonic behavior in the speed of sound, suggesting that such equations of state may be more prone to supporting hybrid stars with quark cores. 

Following this important observation, in the present paper we demonstrate that by using a wide range of different effective dilaton profiles which are also regular in the IR (contrary to \cite{BitaghsirFadafan:2019ofb}) one can achieve stable compact stars as massive as 2 solar masses that include quark cores. A scan over the space of parameters reveals a set of allowed EoS that are consistent with the constraints from \cite{Annala:2017llu}, some of which support quark stars. The improved model also supports stable wrapped D5-branes, which one would in principle identify with nucleons in the dual gauge theory. We show however that (at least in the homogeneous approximation) the resulting nuclear EoS are in conflict with phenomenology and cannot be used in realistic simulations of neutron stars, necessitating the use of the Hebeler-et.al EFT EoS \cite{Hebeler:2013nza} for the baryon phase.
This model is not able to give a definite answer as to whether stable quark stars exist in nature or not, since that depends on the parameter values chosen. It does however show that such quark stars can \textit{in principle} be obtained from holographic models which allows one to explore the phenomenology of such objects.  We find stable quark stars with maximum masses of up to $2.17 M_\odot$ whose tidal deformability drops rapidly upon the formation of quark cores.

The rest of the article is organized as follows. In Sec. \ref{SectionII} we discuss in detail the holographic model we use for the quark matter phase, specifically the dilaton profile responsible for the chiral symmetry breaking mechanism. We also comment on an attempt to include a baryonic phase within the model using D5-branes. In Sec. \ref{SectionIII} we show the compact star phenomenology obtained from the model in the previous section. We solve the TOV equations for an ensemble of EoS to find the mass vs. radius curves showing stable massive configurations of quark stars. We also compute the tidal deformability and compare our results to the LIGO/Virgo data. Finally, in Sec. \ref{SectionIV} we summarize and conclude.

\section{Holographic Model, Quark and Baryon phases}\label{SectionII}

In this section we present the model used throughout this work and discuss the different phases it predicts. {In section \ref{quarkphase} we discuss the action, equations of motion and solutions for the quark phase. In section \ref{baryons} holographic baryons as wrapped D5-branes are analyzed. Due to their unrealistic pressures they are discarded in later sections in favor of the phenomenological Hebeler-et.al EoS \cite{Hebeler:2013nza}.}

\subsection{Set-up and quark phase}\label{quarkphase}
 
{The starting point of the model is the D3/D7-brane configuration \cite{Karch:2002sh} dual to the $\mathcal{N}=4$ $SU(N_c)$ SYM with $N_f$ matter hypermultiplets in the fundamental representation, which has $\mathcal{N}=2$ supersymmetry}. The dual geometry in the Einstein frame is
\beq ds^2 =  {w^2 \over R_{\text{AdS}}^2} (- g_t dt^2 + g_x d\vec{x}^2) + {R_{\text{AdS}}^2 \over w^2} dw_6^2.  \eeq
Here $g_t$ and $g_x$ are metric functions {that can in principle accommodate for nonzero temperatures but for the purposes relevant to this paper they can both be set equal to $1$}. The AdS radius $R_{\text{AdS}}$ obeys the familiar relation
\bea
\frac{R_{\text{AdS}}^4}{\alpha'^2}=4 \pi g_s N_c = \lambda_t,
\eea
{where $\alpha'=l_s^2$, with $l_s$ the string length}, $g_s$ being the string coupling and $\lambda_t$ being the t'Hooft coupling.
Considering D7-branes in the D3 background, corresponds to adding $N_f$ quarks to the system. {Using for} $\mathbb{R}^6$ {the coordinate system}  
\bea
\label{eq:R6}
dw_6^2=d\rho^2+\rho^2 d\Omega_3^2+ d\chi^2+\chi^2 d\varphi^2
\eea
{allows one to parametrise }the D7-brane {embedding by a single function $\chi(\rho)$ }.
Its world volume can then be described by $(t, \vec{x},\rho, \Omega_3)$ coordinates as long as $\chi(\rho)$ is monotonic.
The coordinates $\chi$ and $\rho$ are related to $w$ via 
\bea
w=\sqrt{\rho^2+\chi(\rho)^2}.
\eea
{T}he dilaton profile  \cite{Evans:2012cx}
\bea 
\label{eq:dil}
	e^{\phi}=A+1-A ~ \tanh \bigg[\kappa\bigg(\frac{ w-\lambda}{R_{\text{AdS}}}\bigg)\bigg]
\eea
interpolates smoothly and monotonically from $1$ in the UV to a finite adjustable value in the IR. The parameters $A,\lambda, \kappa$ determine the precise approach to the IR value. The only scale at zero temperature that breaks the conformal symmetry in the system is $\lambda$ {while $A$ controls the IR value of the dilaton, and $\kappa$ the steepness}.

The DBI action for $N_f$ D7-branes reads
\begin{equation}
    S_{D7}=-\frac{T_{D7}}{g_s} \int d^{7+1}x e^{-\phi} \text{Tr}\sqrt{-\det ( G_{\alpha \beta }+2 \pi \alpha' F_{\alpha \beta})},
\end{equation}
with $G_{\alpha \beta }$ being the induced metric and $T_{D7}= (2 \pi)^{-7} \alpha'^{-4}$. {It} describes the degrees of freedom on the D7-brane and using the symmetries {it} can be written as
\bea
S_{D7}=-\bar{T}_{D7}~\int~ dt d^3x d\rho ~\call_{D7},
\eea
where 
\begin{equation}
\label{eq:D7lagr}
  \mathcal{L}_{D7}=   \rho^3 g_x^{\frac{3}{2}} e^{{\phi}} \sqrt{(1+\chi'^2)g_t- e^{- {\phi}} (\partial_{\rho}A_t)^2 }.
\end{equation}
{A}bove  we have {already transformed}  to dimensionless $\rho, \chi$ using the AdS radius and we have switched to a rescaled gauge field $A \frac{2 \pi \alpha'}{R_{\text{AdS}}} =\tilde{A}$ and subsequently dropped the tilde. The constant in front of the action reads
\begin{equation}
\label{eq:T7bar}
    \bar{T}_7= T_{D7} \frac{N_f}{g_s} V(S^3) R_{\text{AdS}}^4= \frac{N_f N_c}{(2 \pi)^4}\frac{\lambda_t}{R_{\text{AdS}}^4}.
\end{equation}
{where $V(S^3)=2\pi^2$ is the volume of the round 3-sphere.}
The coordinate $A_t$ is cyclic {and its}  conserved charge is defined by
\bea \label{quarkdensity}
{d}=\frac{\del \call_{D7}}{\del {A}'_t}= \frac{\rho^3 e^{\phi}\partial_{\rho}A_t}{\sqrt{1+ \chi'^2- e^{-\phi} (\partial_{\rho}A_t)^2}},
\eea
which can be used to express $\partial_{\rho} A_t$ in terms of the constant density $d$.
Similarly the equations of motion for $\chi$ are derived from the Lagrangian \eqref{eq:D7lagr}.

Solving the equations of motion for $\chi$ and $A_t$ subject to
\begin{align}
\chi(\infty)= m, \nn \\
{A}_t(\infty)= {\mu},
\end{align}
one obtains the configurations that are realized in the grand canonical ensemble at fixed chemical potential\footnote{Note that due to the rescaling of the gauge field the quantity $\mu$ is not the physical quark chemical potential.} $\mu$. The parameter $m$ describes an analogue of a quark mass in the dual 4d field theory {and is set to zero in this paper}. 

As the equations of motion are second order,  {one} need{s} one more integration constant for $\chi$ and $A_t$ each \cite{Babington:2003vm} 
\bea
\chi(\rho)= m+ \frac{c}{\rho^2}+...\nn \\ 
{A}_t(\rho)= {\mu}-\frac{{d}}{2 \rho^2}+...
\eea
The constant $c$ measures spontaneous chiral symmetry breaking and $d$ is the density of quarks (or baryons after dividing by $N_c$). {Their values are determined dynamically by the boundary conditions in the IR. These are best explained by turning on a small nonzero temperature resulting in a non-vanishing black hole horizon $w_H$}.

If $\sqrt{\rho^2+\chi^2(\rho)}> w_H$ for all $\rho \ge0$ then $\chi'(0)=0$ is necessary together with $d=0$.  If there is a value $\rho_c$ for which $\sqrt{\rho_c^2+\chi^2(\rho_c)}= w_H$ then the $(\rho,\chi(\rho))$ curve in the $(\rho,\chi)$ plane has to intersect $\sqrt{\rho^2+\chi^2}= w_H$ orthogonally and in addition $A_t(\rho_c)=0$.
This is because in the Euclidean version of the black hole geometry, the time coordinate is an angular variable and the black hole horizon is just the origin of the coordinate system. {The zero temperature case is recovered by letting $w_H \rightarrow0$}
\begin{figure}
	\includegraphics[width=1\linewidth]{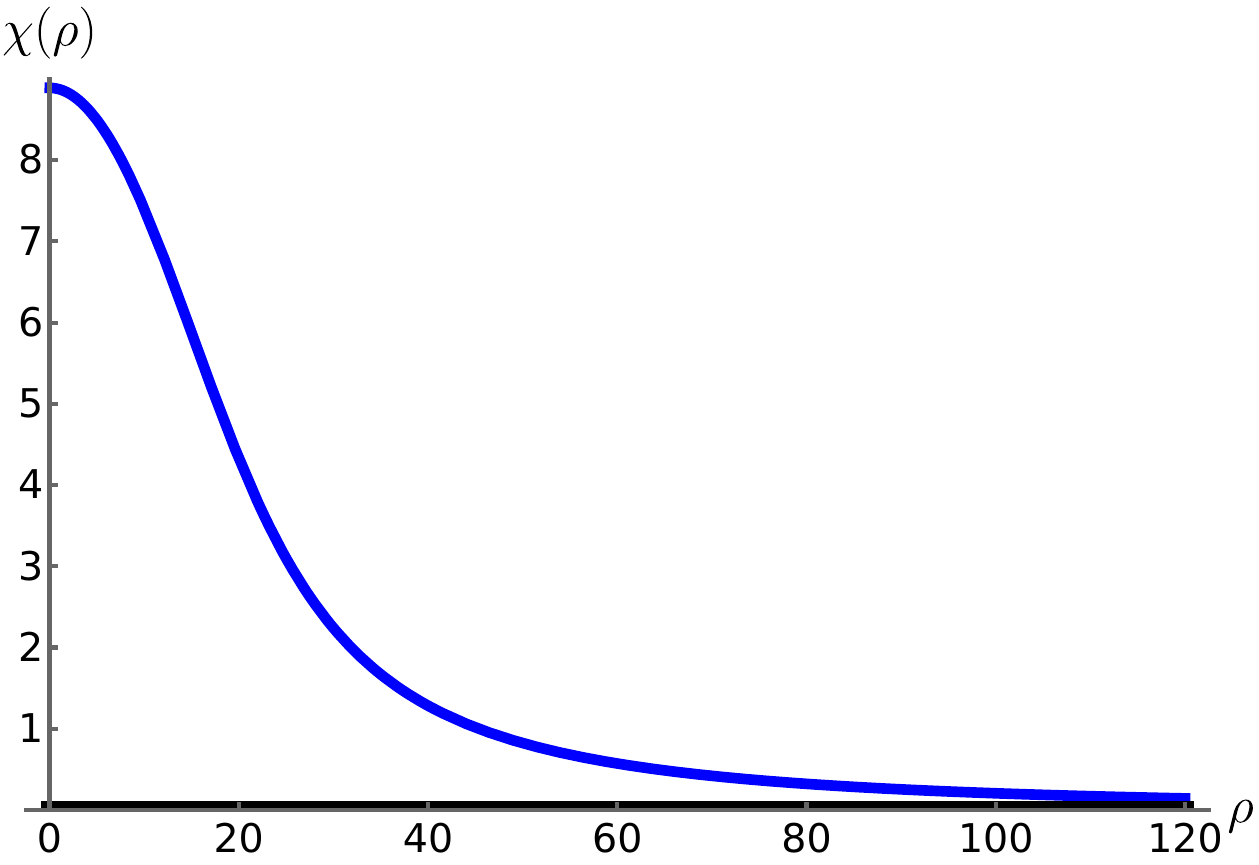}  
	\includegraphics[width=1\linewidth]{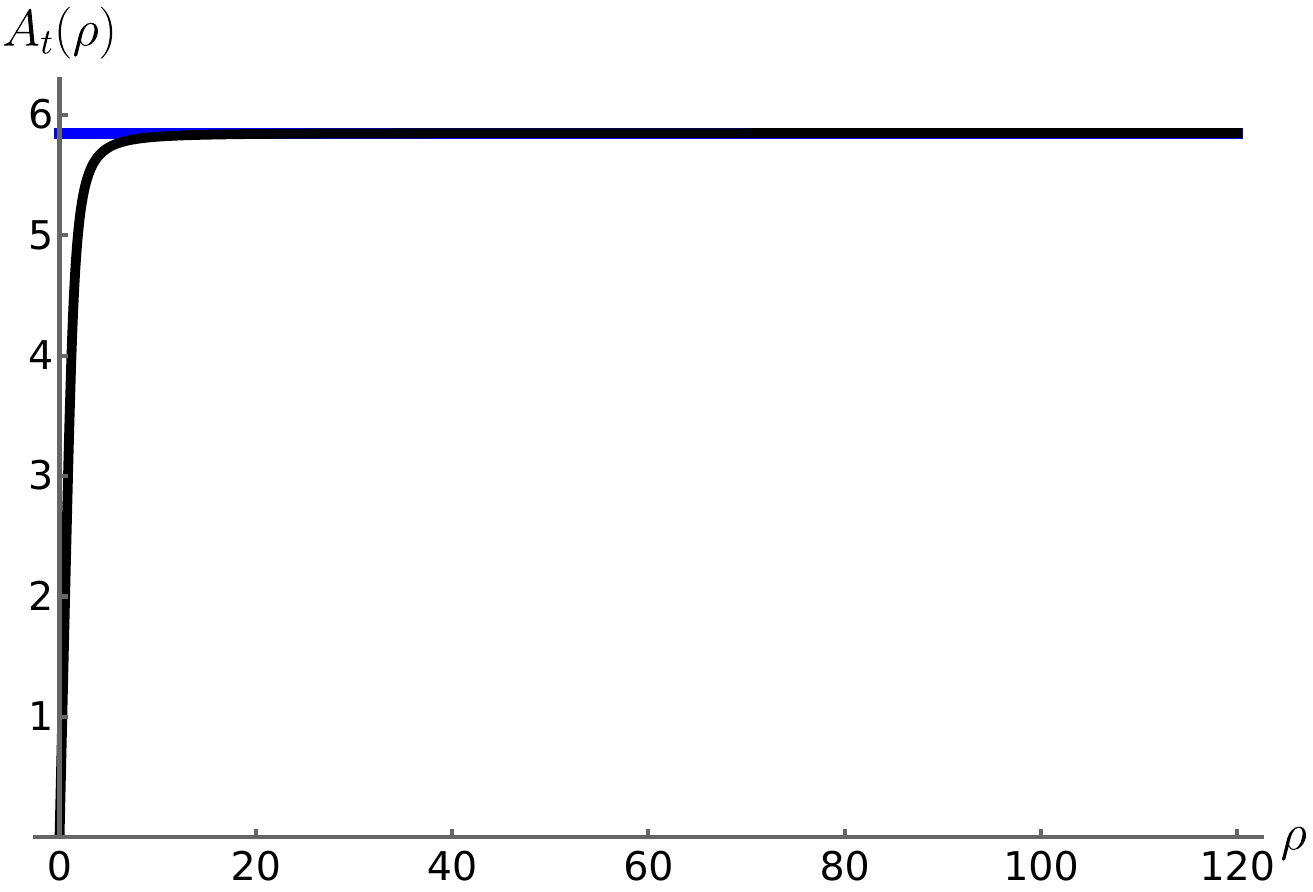} 
	\caption{\footnotesize{ The {upper} plot shows the embedding function $\chi(\rho)$ in the Minkowski {(or vacuum)} phase (blue) and the quark phase (black) for a generic value of $\mu$ at zero temperature. The {lower} plot shows the corresponding gauge potentials $A_t(\rho)$}. {The Minkowski phase has a chiral symmetry breaking vacuum but zero baryon density, while the quark phase restores chiral symmetry and has a nonzero $d$. All quantities are dimensionless and computed for a generic point in parameter space.}}
	\label{fig:chAt}
\end{figure}

{The holographic dictionary equates the on-shell action (stripped off $\int d^3x dt$) $S_{D7}^{os}$ with the grand canonical potential density $   \Omega$, which is related to the pressure via
\begin{align}
    P(\mu)=-\Omega(\mu).
\end{align}
}

We call the phase with zero density (for all $\mu$) the Minkowski {(or vacuum)} phase and its embedding and gauge potential are shown in blue in figure \ref{fig:chAt}. Since $\frac{d}{d \mu} S_{D7}\sim d$, the action and hence the grand canonical potential density are constant as a variable of $\mu$ {in this phase}.
The phase for which $\sqrt{\rho^2+\chi^2}$ reaches $ w_H=0$ is called the quark phase {which is depicted in black in figure \ref{fig:chAt}}. At large $\mu$ this phase  dominate{s} and at asymptotically large densities reproduces the QCD EoS $P(\mu)\sim \mu^4$ . 

\begin{figure}
    \centering
    \includegraphics[width=1 \linewidth]{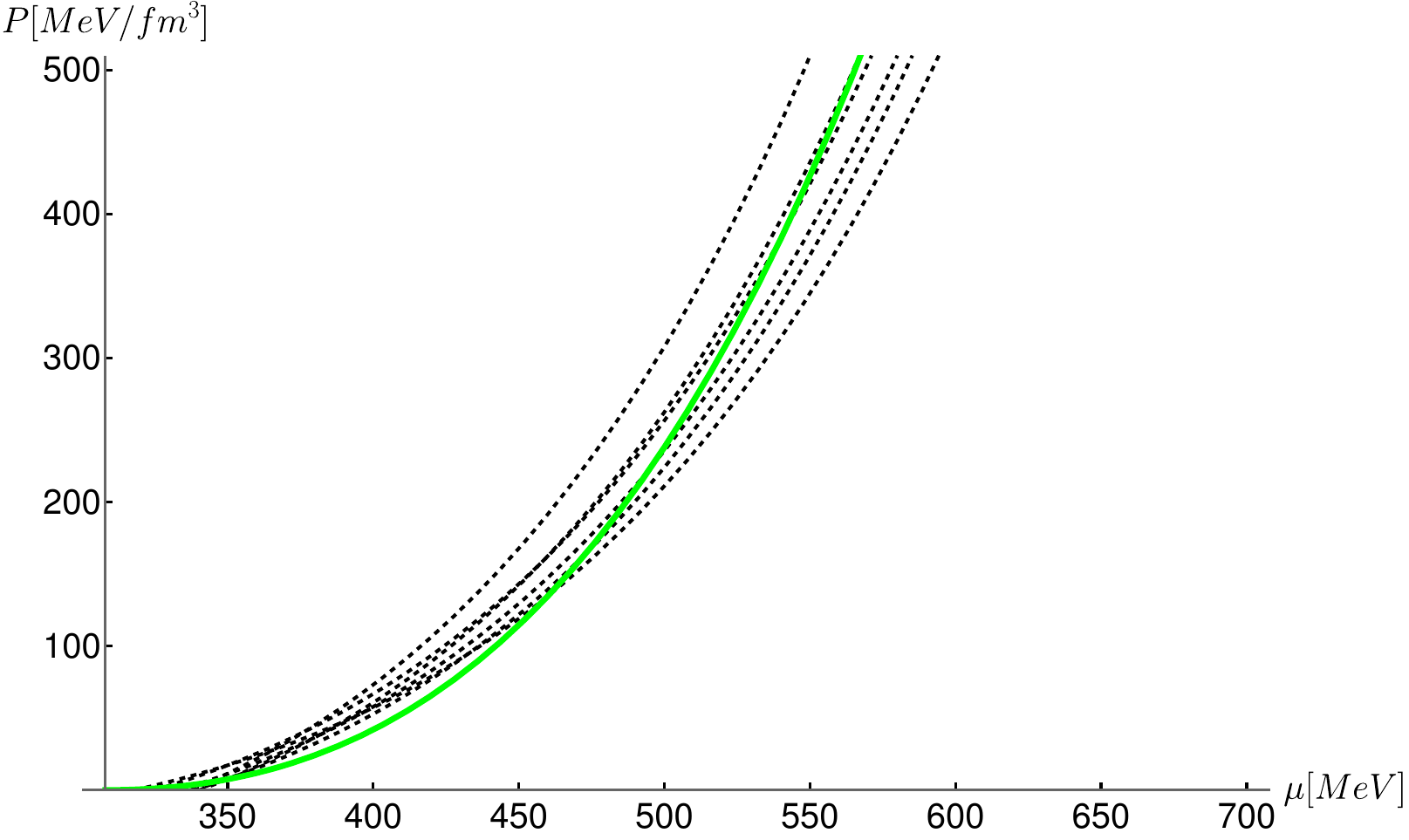}
   
    \caption{Pressure $P$ as a function of the chemical potential for the quark phase defined with dilaton profile \eqref{eq:dil} and range of parameters $\lambda_t=1.9...3$, $R_{\text{AdS}}= 0.015...0.02$ MeV$^{-1}$ (dashed black). The green curve corresponds to the basic D3/D7 model of \cite{Hoyos:2016zke} . The reduced steepness of {some of the} the dashed black curves allows for a smoother transition from baryonic matter to quark matter ultimately enabling stars with quark cores when using the stiff Hebeler-et.al EoS for nucleon matter.}
    \label{fig:3}
\end{figure}

In figure \ref{fig:3} we plot the pressure {in the quark phase} as a function of the chemical potential {for} a range of parameters given by $\lambda_t=\{1.9, ..., 3\}$, $R_{\text{AdS}}=\{0.015,...,0.02\}$ MeV$^{-1}$ {(we explain our choices for the other parameters in section \ref{sec:pars}) and compare it to the basic D3/D7 model with trivial dilaton of \cite{Hoyos:2016zke} which is also reproduced in appendix \ref{appendixA} }. {T}he reduced steepness at smaller $\mu$ of {some of} the {new}  curves {(with non-trivial dilaton)} allows for a smoother transition from baryonic matter to quark matter which leads to stars with quark cores.

\subsection{Baryons as wrapped D5 branes}
\label{baryons}
{In this subsection we analyze the baryons intrinsic to the model. As shown in \cite{Evans:2012cx} a non-trivial dilaton like \eqref{eq:dil} can stabilize D5-branes wrapping $S^5$, which have to be interpreted as baryons in the dual gauge theory.}

Due to the nonzero {background} 5-form flux {in AdS} the world volume gauge field of the D5-brane sees an effective charge $N_c$. {T}he spatial world volume is compact {which implies that} the total charge on the D5 brane {needs to vanish} {and therefore} other charge sources {are} need{ed} to neutralize this uniform charge density. These are fundamental strings with one end on the D5 {brane} and one end on the {stack of} D7-branes. This in turn introduces localized charges on the D7-branes, but here the flux can escape to infinity and contribute to the asymptotic value of (the time component of) the D7 gauge field and hence the density $d$.

To study the D5-brane, we parameterize {the} $\mathbb{R}^6$ part of the geometry \eqref{eq:R6} as
\begin{equation} \label{D5Paramerization}
	 dw_6^2 = \dd \xi^2 + \xi^2 \left(\dd\theta^2+ \sin{\theta}^2 \dd\Omega_4^2\right) \,,
\end{equation}
The D5-brane will be {described by} $(t,\theta, \Omega_4)$ {coordinates allowing for}  a non-trivial profile\footnote{The coordinate $\xi$ is actually the same as the coordinate $w$ introduced earlier. Since the D7-brane embedding is already described by a function $w(\rho)$, it is advantageous to nonetheless introduce $\xi$ and describe the D5-brane embedding by the function $\xi(\theta)$}  $\xi (\theta)$ where $\theta \in [0,\pi]$. {The brane configurations and relevant coordinate systems in this model are summarized in table} \ref{tableDbranes}.

\begin{table}[h!]
\begin{tabular}{|c|c|c|c|c|c|c|c|c|}
\hline & \multicolumn{5}{|c|}{ AdS-Black hole } & \multicolumn{3}{|c|}{$S^5$} \\
\hline coordinate & $t$ & $x^1$ & $x^2$ & $x^3$ & $ w$ & \multicolumn{3}{|c|}{$S^5$} \\
\hline Background D3 & $\bullet$ & $\bullet$ & $\bullet$ & $\bullet$& & \multicolumn{3}{|c|}{}   \\
\hline & & & & & \multicolumn{3}{|c|}{$\mathbb{R}^4$} & ${\mathbb{R}}^2$ \\
\hline coordinate & $t$ & $x^1$ & $x^2$ & $x^3$ & $\rho$ & \multicolumn{2}{|c|}{$S^3$} & ${\mathbb{R}}^2$ \\
\hline Flavor D7 & $\bullet$ & $\bullet$ & $\bullet$ & $\bullet$ & $\bullet$ & \multicolumn{2}{|c|}{$\bullet$} & \\
\hline & & & & & $\mathbb{R}$ & \multicolumn{3}{|c|}{$S^5$} \\
\hline coordinate & $t$ & $x^1$ & $x^2$ & $x^3$ & $\xi$ & $\theta$ & \multicolumn{2}{|c|}{$S^4$} \\
\hline Baryon vertex D5 & $\bullet$ & & & & &$\bullet$ & \multicolumn{2}{|c|}{$\bullet$} \\
\hline
\end{tabular}
\caption{{This table indicates in which directions and coordinates the various branes in this setup are extended. A dot indicates that the brane stretches in this direction.} }
\label{tableDbranes}
\end{table}

{Computing} the embedding of a single localized (say near $\vec{x}=\vec{x_0} \in \mathbb{R}^3$) D5-brane, {obeying the classical equations of motion}, is {a} very {challenging problem}. {Multi-localized D5 solutions are even more difficult to find\footnote{Additionally for any given baryon number, there should be multiple saddle points that contribute. One can always include any given number of baryons-antibaryon pairs without changing the total baryon number.}. Only at high densities would one expect to be able to approximate such configurations by a single smeared D5-brane.
Localized D5-branes are beyond the scope of this paper and we proceed by using the smeared brane approximation tentatively \textit{for all densities}. {As we will show, particularly at low densities there will be some  tension between the prediction from the smeared brane approximation and experimental data, pointing towards the need to localize the D5 branes.}}

The action of the smeared branes is a sum of DBI and topological terms and is given by
 \begin{eqnarray}\label{D5action}
 	\frac{S_{D5}}{n_{D5}}&=&-\frac{T_{D5}}{g_s} \int{\dd^{5+1} x ~ e^{-\phi}
 		\sqrt{-\text{det}(\mathcal{G}_{\alpha\beta}+2 \pi \alpha'\mathcal{F}_{\alpha b})}}\nn\\ &&+ 2 \pi \alpha' T_{D5}\int{ \cala \wedge G_{(5)}}
 \end{eqnarray}
 where $\mathcal{G}_{\alpha\beta}$ is the pullback of the string frame metric to the D5  brane and $n_{D5}= \int d^3x \rho_{D5}$ is the total number of D5-branes\footnote{The density $\rho_{D5}$ is related to the density $d$ in equation \eqref{eq:chargecons} below. }. The tension of the D5-brane is $T_{D5}= (2\pi)^{-5} \alpha'^{-3}$ and the world volume gauge field of
 the D5-brane $\cala$ (with field strength $\mathcal{F}$) is sourced by the five form $G_{(5)}$ of the background geometry.

 The $G_{(5)}$ flux that the background induces is given by
 \begin{align}
     G_{(5)}=2 \kappa_0^2 T_{D3} N_c \frac{d\Omega_5}{V(S^5)}=\kappa_0^2 T_{D3} N_c \frac{d\theta (\sin \theta)^4 d\Omega_4}{V(S^5)},
 \end{align}
 where $2\kappa_0^2=(2\pi)^7 \alpha'^{4}$, $T_{D3} = {(2\pi)^{-3}} {\alpha'^{-2}}$ and the integration measures on {(and volumes of)} $S^5, S^4$ are defined in the standard way.

 In \eqref{D5action} the dilaton reads
 \begin{equation}\label{eq:dil2}
e^\phi=A+1-A \tanh \left[\kappa\left(\frac{\xi(\theta)-\lambda}{R_{\text{AdS}}}\right)\right].
\end{equation}

 {We choose to rescale the dimensionful quantities} $\xi= R_{\text{AdS}} \tilde{\xi}$, $\lambda= R_{\text{AdS}} \tilde{\lambda}$ and ${2 \pi \alpha'}\cala /{R_{\text{AdS}}} =\tilde{\cala} $, where $R_{\text{AdS}}$ is the AdS radius. Through an abuse of notation we will {also} drop the tilde on $\tilde{\xi},\;\tilde{\mathcal{A}}$ henceforth.
 
  {Inserting} $\xi=\xi(\theta)$ and $\cala_t(\theta)$ {into} the D5-brane action gives 
 \begin{eqnarray}
	S_{D5}&=&-\bar{T}_{D5} \rho_{D5} \int dt d\theta \call_{D5}, 
\end{eqnarray}
with
\begin{align}
     \mathcal{L}_{D5}&=  e^{\frac{{\phi}}{2}} \sin^4 \theta \sqrt{g_t(\xi^2+\xi'^2)- e^{-{\phi}}(\partial_{\theta}\cala_t)^2 } d\theta \;
    - 4 \cala_t \sin^4 \theta 
\end{align}
and $\bar{T}_{D5}$ collects all the prefactors except the D5 density, which is explicitly displayed. In the grand canonical ensemble which we consider later, the chemical potential is specified and the D5 density needs to be solved for. It is a dynamical variable just like the embedding function or the gauge field.

The conjugate momentum of ${\cala}_t(\theta)$ is defined by $D={ \partial \call_{D5} }/{\partial {\cala}_t'(\theta)}$ as
\begin{eqnarray}
D(\theta)=\frac{\cala'_te^{-\frac{{\phi}}{2}}\sin^4 \theta  }{\sqrt{g_t(\xi^2+\xi'^2)- e^{-{\phi}}(\cala_t')^2 }}.
\end{eqnarray}

The equation of motion of ${\cala_t}(\theta)$ reads 
\bea
\label{eq:Doftheta}
\del_{\theta} D(\theta)= 4\sin^4 \theta .
\eea
This equation describes the worldvolume gauge fields correctly at all points $\theta$ except where the fundamental strings are attached. They would contribute additional delta terms to \eqref{eq:Doftheta}.

The solution of this equation is
\bea
D(\theta)= \frac{3 \theta }{2} -\sin( 2 \theta )+\frac{1}{8} \sin( 4 \theta).
\eea

\begin{figure}[h] 
	\includegraphics[width=8.5cm]{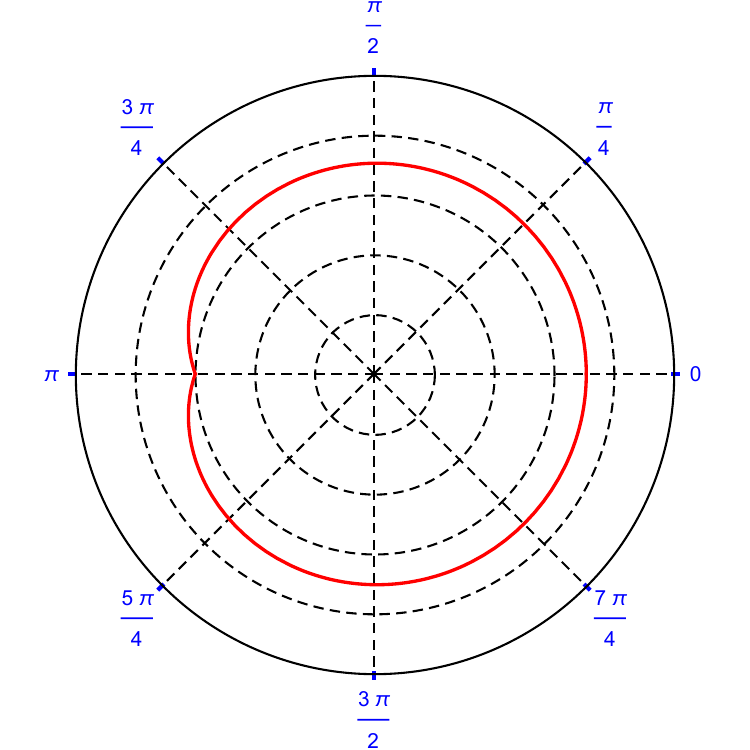}  
		\caption{\footnotesize{ { The baryon vertex for $A=10,\tilde{\lambda}=1.715,\kappa=1$. All fundamental strings emerge from the pole at $\theta=\pi.$}}}
        \label{fig:baremb}
\end{figure}

The integration constant in this expression is zero which implies that the strings emerges only from one of the poles of the baryon vertex which is located at $\theta=\pi$. The equation of motion for $\xi$ has to be solved numerically with initial conditions $\xi(0)=\xi_0,
$ and $ \xi'(0)=0$. The non-smoothness of the embedding at $\theta= \pi$ is caused by the {fundamental} strings which attach there, see figure \ref{fig:baremb}.


In order to understand the system as a whole we have to also include the D7-branes and the strings connecting the D5 and D7-branes.

The total action density (per unit 3-volume) for the joint system of strings, the smeared D5-brane and the $N_f$ D7-branes reads \footnote{Both gauge fields have been rescaled and are dimensionless here. The D7-brane coordinates are dimensionless as well.}
\begin{align}
\label{eq:acttot}
-S=&\rho_{D5} \overline{T_{D5}} \int d\theta \mathcal{L}_{D5} +\overline{T_{D7}} \int d \rho \mathcal{L}_{D7} \nonumber \\
&+\frac{\rho_{D5} N_c R_{\text{AdS}}}{2 \pi \alpha'}\cala_t(\theta=\pi)-\frac{\rho_{D5} N_c R_{\text{AdS}}}{2 \pi \alpha'}A_t(\rho=0) 
\end{align}
which reduces to the grand potential density when on-shell configurations for the fields are inserted.
Energetic arguments suggest that the strings shrink to zero length \cite{Gwak:2012ht} and hence the area of the fundamental strings  vanishes and all that remains are the couplings to the world volume gauge fields in line 2 {together with} the boundary condition $\xi(\pi)= \chi(0)$.

Every variation of \eqref{eq:acttot} has to vanish at a solution to the equations of motion satisfying the proper boundary conditions. Varying $\xi, \chi$ and the two gauge fields away from $\rho=0$ and $\theta= \pi$ gives equations of motion already stated above.

{Demanding $\delta S=0$ for nonzero $\delta \cala_t (\pi)$ and $\delta A_t(0)$ results in equations relating the charges of the different branes}

\begin{align}
     \overline{T_{D5}}(2 \pi \alpha')D(\pi)=N_c R_{\text{AdS}}, \nonumber \\
     \overline{T_{D7}}(2 \pi \alpha')d= \rho_{D5} N_c R_{\text{AdS}},
     \label{eq:chargecons}
\end{align}

Similarly for variations of $\chi, \xi$ that are nonzero at the boundary (but still obey $\delta \xi(\pi)= \delta \chi(0)$) one obtains the force balance condition
\begin{align}
     \frac{\partial \mathcal{L}_{D7}}{\partial \chi'}\delta \chi|_{\rho=0} =\frac{2}{3 \pi} d \frac{\partial \mathcal{L}_{D5}}{\partial \xi'} \delta \xi|_{\theta= \pi},
\end{align}
which {can be simplified to}
\begin{align}
    \frac{\xi'(\pi)}{\xi(\pi)}= \chi'(0).
\end{align}
Finally as mentioned before, $\rho_{D5}$ is a dynamical quantity and we also need to vary it.
The resulting equation is
\begin{equation}
\label{eq:gaugeinitialrel}
      \frac{2}{3 \pi }\int d\theta \mathcal{L}_{D5}-A_t(0) +\cala_t(\pi)=0.
\end{equation}
which we solve by setting $\cala_t(\pi)=0$ and $A_t(0)= \frac{2}{3 \pi }\int d\theta \mathcal{L}_{D5}$\footnote{Any other choice is equivalent.}.

The strategy for solving this system of equations numerically is now as follows: for a given value of $d$ (or by \eqref{eq:chargecons} of $\rho_{D5}$) we solve for $\xi(\pi)$ as a function of the initial value $\xi(0)=\xi_0$. We then use the matching condition $\chi(0)= \xi(\pi)$ and the force balance condition to solve for $\chi,A$ as a function of $\xi_0$. This $\xi_0$ is then tuned such that $\chi(\infty)=0$.
The {(rescaled)} chemical potential is then given by
\begin{align}
\label{eq:chempot}
      \mu =\frac{2}{3 \pi }\int d\theta \mathcal{L}^{os}_5+\int_0^{\infty} \sqrt{\frac{d^2(1+\chi'^2)g_t}{\rho^6 g_x^3+d^2e^{-{\phi}}}}.
\end{align}
with on-shell D5 lagrangian
 \begin{align}
     \mathcal{L}^{os}_5=e^{\frac{{\phi}}{2}}\sqrt{(\xi^2+\xi'^2)g_t}\sqrt{\sin^8 \theta +D(\theta)^2}.
 \end{align}
{T}he on-shell action density is given exclusively by the D7 piece
\begin{align}
\label{eq:grandpot}
    \Omega=-\overline{T_{D7}}\int d \rho \mathcal{L}_{D7} .
\end{align}
{The localized terms in \eqref{eq:acttot} cancel the D5-brane piece upon using \eqref{eq:gaugeinitialrel}.}
Both $\mu$ and $\Omega$ are {thus} parametrized by $d$ and thus can be used to compute $\Omega(\mu)$.
The on shell action {still} requires holographic renormalisation \cite{Henningson:1998gx}, which is {effectively} implemented by integrating only up to $\rho_c$, subtracting $\bar{T_{D7}} \frac{\rho_c^4}{4}$ from \eqref{eq:grandpot} {and taking $\rho_c\rightarrow \infty$}.

{The relation between the physical chemical quark potential and the above rescaled $\mu$ is}
\begin{align}
    \mu= \mu_{\text{phys}} \frac{2 \pi \alpha'}{R_{\text{AdS}}}= \mu_{\text{phys}}\frac{2 \pi R_{\text{AdS}}}{\sqrt{\lambda_t}}.
\end{align}

{The rescaled D7-brane tension} $\bar{T}_7$ {also} depends on $R_{\text{AdS}},\lambda_t$ {which}  can be chosen to match with {experimental} data or with UV QCD. 
There will be a (dimensionless) value $\mu_{\text{min}}$ of $\mu$ such that the grand potential of the D5 phase is the same as that of the Minkowski phase. This {point} marks the phase transition from the vacuum to a phase with baryons and is in real QCD roughly at $\mu_{\text{phys}}=308$ MeV. Hence we choose
$R_{\text{AdS}}=\frac{\sqrt{\lambda_t}}{2 \pi} \frac{\mu_{\text{min}}}{308}$ MeV$^{-1}$.
The t'Hooft coupling $\lambda_t$ is a free parameter and can be chosen for example such that the quark phase reproduces the UV QCD result at high chemical potentials.

Let us briefly comment on our use of $R_{\text{AdS}}$ as a fitting parameter. We have chosen to express dimensionful quantities in terms of $R_{\text{AdS}}$ simply because of convenience. In the holographic model, the stringy modes still decouple and quantities are independent of the string scale $\alpha'$. 

As mentioned before, the non-trivial shape of the dilaton is a key ingredient in getting stable D5-branes. With a constant dilaton, it is energetically favorable to completely shrink the D5-branes, whilst {the} profile \eqref{eq:dil} stabilises them against collapse. {As already shown in \cite{Evans:2012cx} the resulting phase diagram captures important aspects of QCD qualitatively. The onset phase transition to a baryonic phase at $\mu_{\text{phys}}\sim 308 $ MeV which subsequently switches over to a quark phase at higher $\mu_{\text{phys}}$. 

The D5 baryonic phase differs however appreciably {from the phenomenological equations of state of \cite{Hebeler:2013nza} as showcased by figure \ref{fig:D5comp}. Particularly just above the transition from vacuum to baryonic matter the D5 EoS compares poorly with the phenomenological equations of state.}
{A}t small $\mu-\mu_{\text{min}}$ one can use \eqref{eq:chempot} and expand it in the density d to get

\begin{align}
    \mu =\mu_{\text{min}} +d^{\frac{1}{3}} \int_0^{\infty}d \rho \sqrt{\frac{1+\chi'(0)^2}{\rho^6+ e^{-\tilde{\phi}(0, \chi(0))}}} =: \mu_{\text{min}} +d^{\frac{1}{3}}a,
\end{align}
{with a $d$-independent coefficient $a$.}
{Together with} $\frac{d}{d \mu}\Omega \sim d(\mu)$ one {then obtains} $\Omega \sim (\mu-\mu_{\text{min}})^4$ close to $\mu_{\text{min}}$ {suggesting a fourth order transition\footnote{The modification of the action and force balance condition by various new parameters does not fix this problem.}}. {It may well be that this rather undesirable behavior stems from the smeared approximation and that solutions with localized D5-branes improve the situation at low density. It would be interesting to study such solutions in the future. The mass of this configuration depends on the parameters of the theory and would provide also an important check of the parameter choice one uses.}

{For the moment however} our main objective is to describe compact stars that are in agreement with current astrophysical observations, {hence in the absence of knowledge about localized D5 solutions we will describe the baryonic phase phenomenologically as in \cite{Hebeler:2013nza}}. The parameter $R_{\text{AdS}}$, which was previously fixed to relate $\mu_{\text{min}}$ to $\mu_{\text{phys}}^{\text{trans}}=308$ MeV, can now be varied.
Combining holographic EoS with phenomenological curves at low $\mu$ is a rather standard procedure in the literature when using it for neutron star phenomenology \cite{Bartolini:2023wis,Tootle:2022pvd,Jarvinen:2021jbd}. Holographic calculations typically work within a smeared approximation which is expected to be inaccurate at low densities. Additionally, at low densities the effects of leptons on the EoS is substantial. It is necessary to impose beta equilibrium and to also understand the symmetry energy, which has has never been computed with localized holographic matter\footnote{Computations for smeared holographic matter were presented in \cite{Bartolini:2022gdf}}. Techniques for extracting EoS including all these effects from holographic models are still under development. Using a phenomenological equation of state (EoS) at low densities is a temporary but necessary workaround in this model to produce realistic neutron stars.

{In the next section we describe our parameter choices and equations of state. We find a region that, together with the stiff phenomenological EoS, produces stable quark stars whose properties we then explore.}

\begin{figure}
    \centering
    \includegraphics[width=1\linewidth]{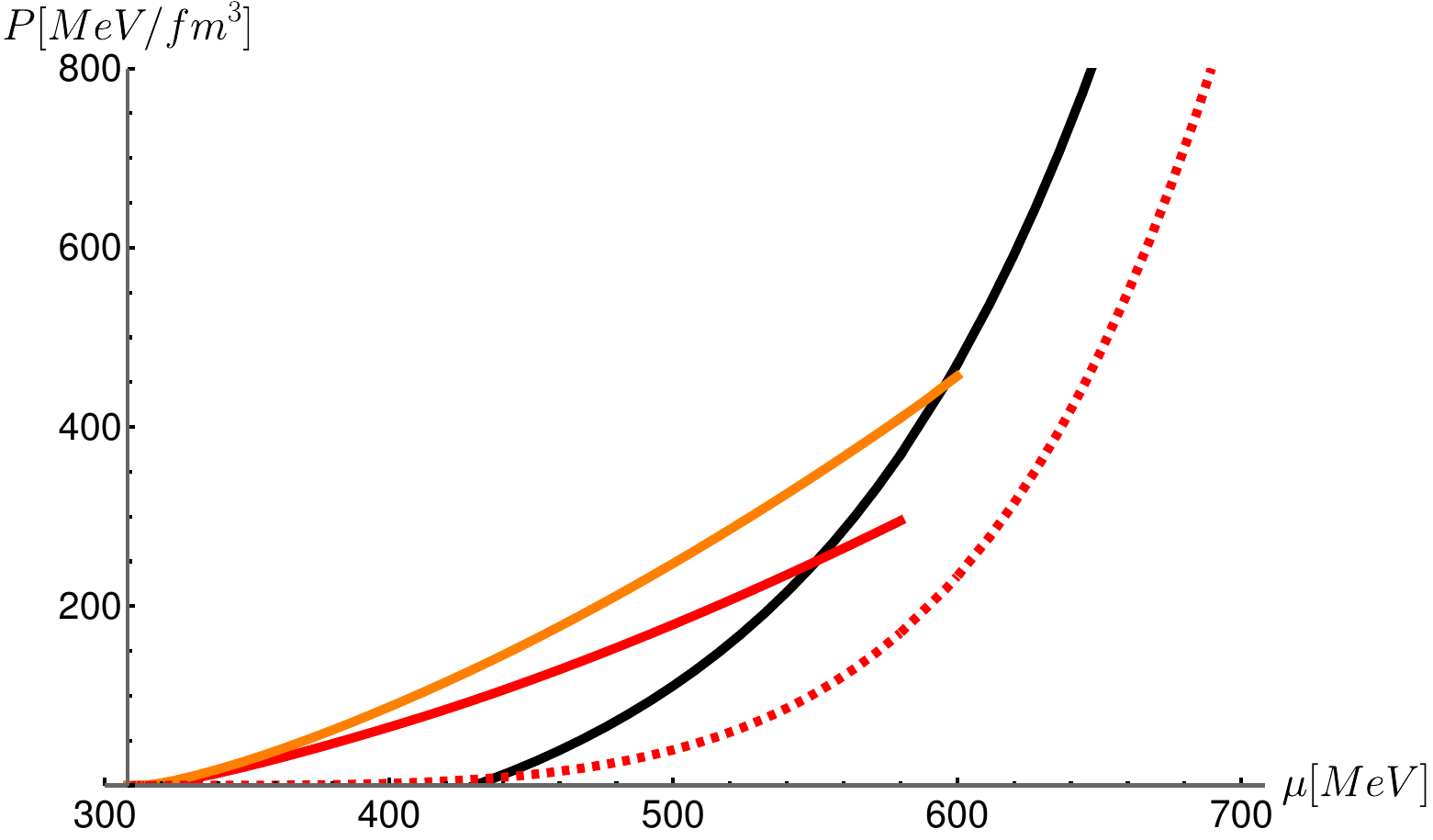}
    \caption{This figure compares the $P(\mu)$ behaviors of the $D5$ baryon phase (dashed red) to the phenomenological medium (in orange) and stiff (in red) phases of \cite{Hebeler:2013nza} (we do not consider the soft EoS). The quark phase is shown in black for the parameter choice $A=10$, $\lambda=1.715$, $\kappa=1$, $\lambda_t=1.9$.}
    \label{fig:D5comp}
\end{figure}

\section{Quark Star Phenomenology}\label{SectionIII}
In this section we will first describe our choices for the free parameters of the theory and subsequently explore the equations of state and the consequences for neutron star observables such as mass, radius and tidal deformability. Let us reiterate, that due to the poor phenomenology of the smeared baryon phase, we resort to the phenomenological Hebeler-et.al for describing the nuclear phase. The quark phase is taken from the holographic model.

\subsection{Parameters}
\label{sec:pars}
For the parameters $A,\tilde{\lambda}$ we adopt the preferred choice of {\cite{Evans:2012cx} of $A=10, \tilde{\lambda}=1.715$. There the parameter $\kappa$ was set to $1$ allowing for the stable D5 solutions discussed in the previous section. As we have seen, the resulting nuclear EoS (in the smeared approximation) is rather unphysical and cannot be used as input for neutron star computations. We find that smaller values of $\kappa$ in general produce weaker first order transitions between baryon and quark matter. We therefore make the choice $\kappa=0.1$. We scan over the values of the two remaining parameters $R_{\text{AdS}},\lambda_t$ and highlighting in our plots 6 sample curves. The region $(R_{\text{AdS}},\lambda_t)=(0.015, ..., 0.02,1.9, ..., 3)$ produces rather stiff EoS with weak first order transitions, some of which weak enough to produce quark stars (those parameter values are mentioned at the end of section \ref{subsec:massradd}). A more extensive exploration of parameter space is left for future work.}

\subsection{Equations of state and constraints}

With the selection of parameters described in the previous section one can readily compute the pressure as a function of $\mu$ and find the transition to baryonic matter. The resulting curves are shown in figure \ref{pressure2} for the stiff (red) and medium (orange) Hebeler-et.al EoS. In the medium case, the transitions lie mostly around $\mu=500$ MeV, whilst for the stiff EoS the phase transitions happen at considerably lower values of the chemical potential, at around $400$ MeV. In the majority of cases (marked in gray), it is not possible to generate stable quark cores, but in the stiff case those with particularly early transitions (marked in black) support quark matter cores as shown in section \ref{subsec:massradd}.

\begin{figure}[h]
    \centering
    \includegraphics[width=1\linewidth]{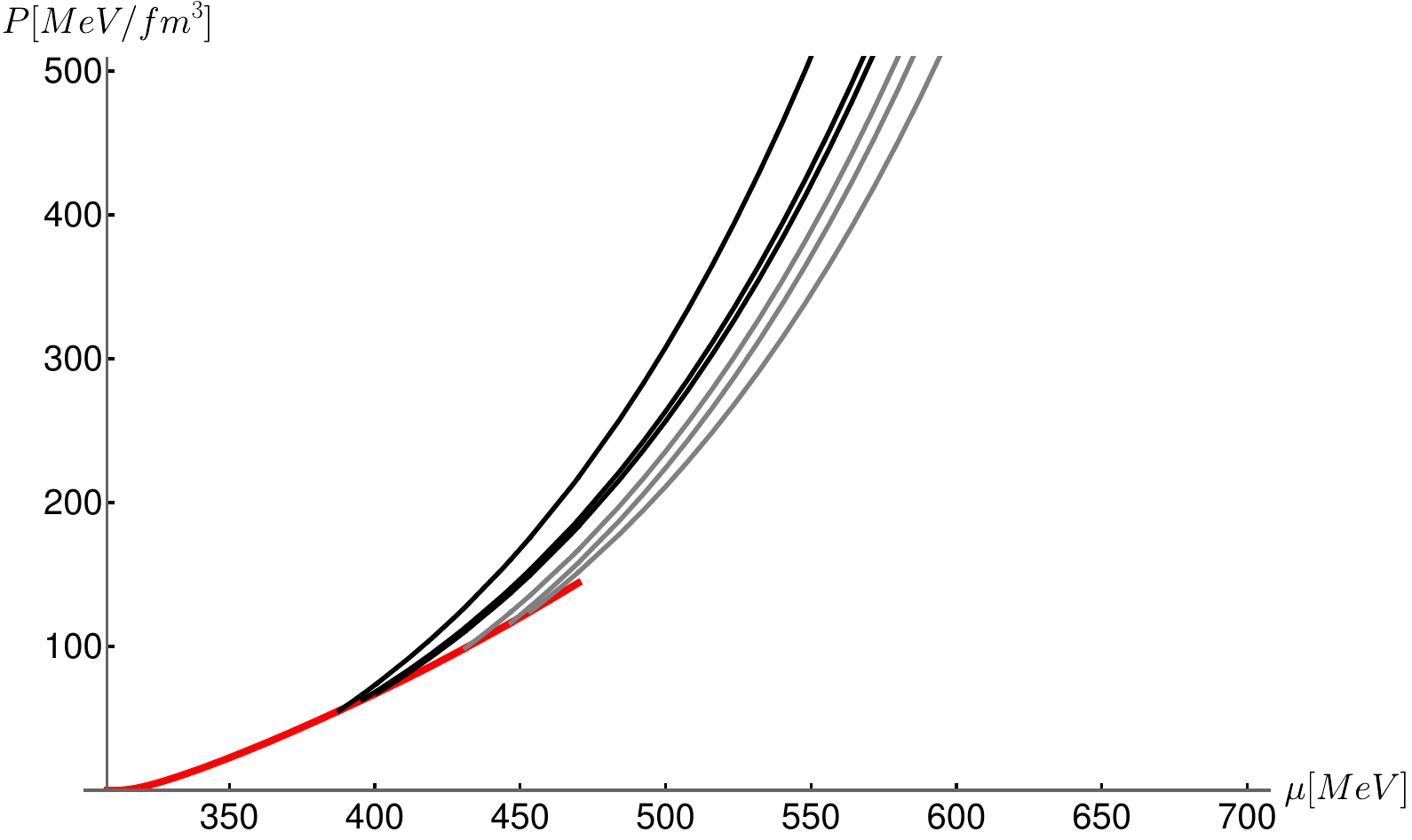} 
    \includegraphics[width=1\linewidth]{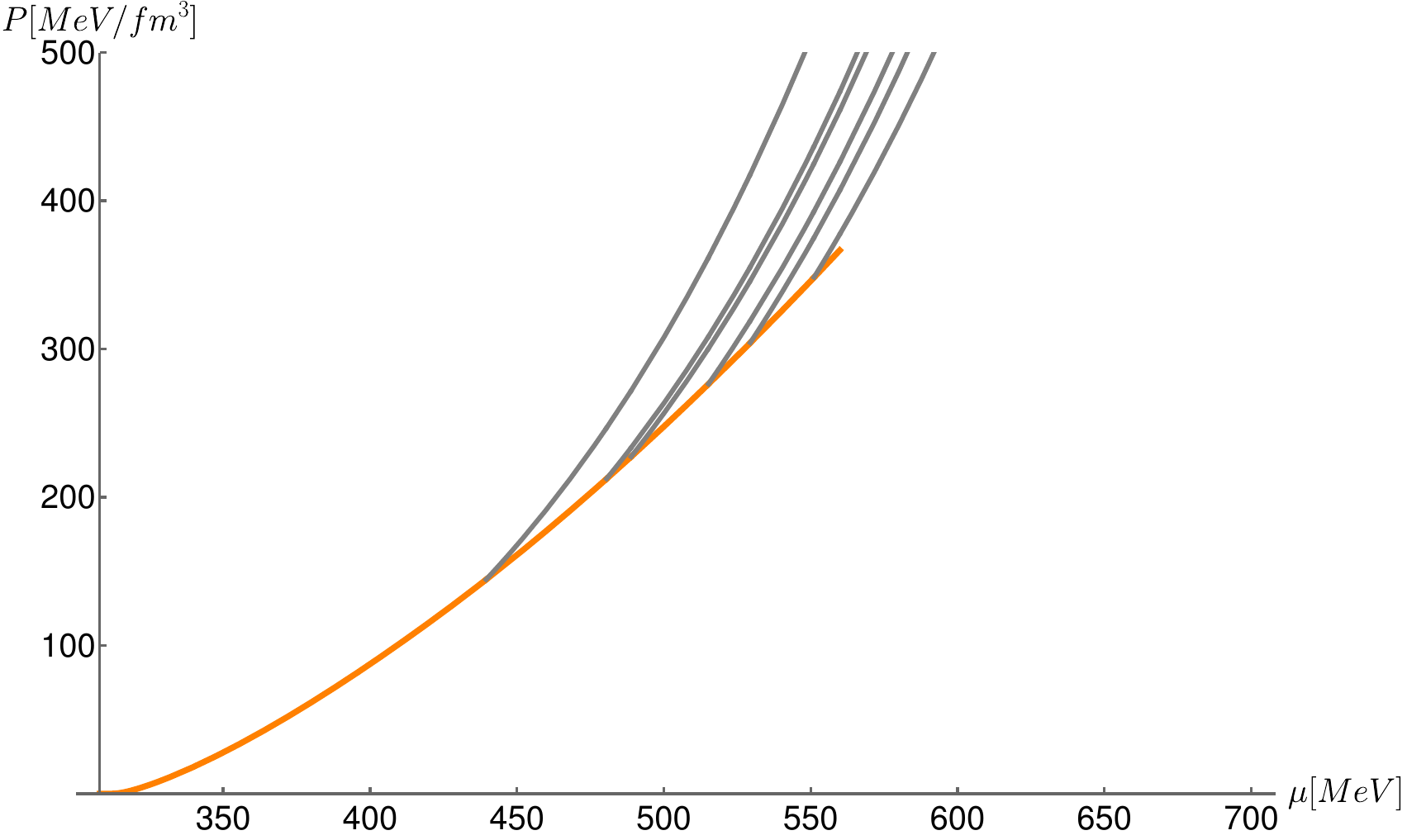} 
    \caption{Pressure as a function of chemical potential for the case of stiff (top) and medium (bottom) baryonic matter of \cite{Hebeler:2013nza}, and quark matter (black and gray) of the holographic model described in section \ref{quarkphase}. The cases colored in black will support stable quark stars while the ones in gray will not. The parameter choices, which are described in section \ref{sec:pars}, are the same in both plots. The plots only differ in the description of the baryonic phase.}
    \label{pressure2}
\end{figure}

One may also study the pressure as a function of energy density $\epsilon$ as shown in figure \ref{banana}. The various quark phases all lie within the gray band, which is obtained in \cite{Annala:2017llu} by considering a large ensemble of equations of state consistent with astrophysical observations (such as $M_{\text{TOV}}>2 M_{\odot}$ and $\Lambda({1.4 M_{\odot}})<580$) and pQCD constraints at large $\epsilon$.
The stiff baryonic phase is clearly quite close to the upper boundary, and at the locations where the phase transitions happen even slightly outside, of this band. From the point of view of \cite{Annala:2017llu} it represents a somewhat extreme EoS, but in our setup it is necessary to have a stiff baryonic EoS to obtain stable quark cores.

\begin{figure}[h]
    \centering
    \includegraphics[width=1\linewidth]{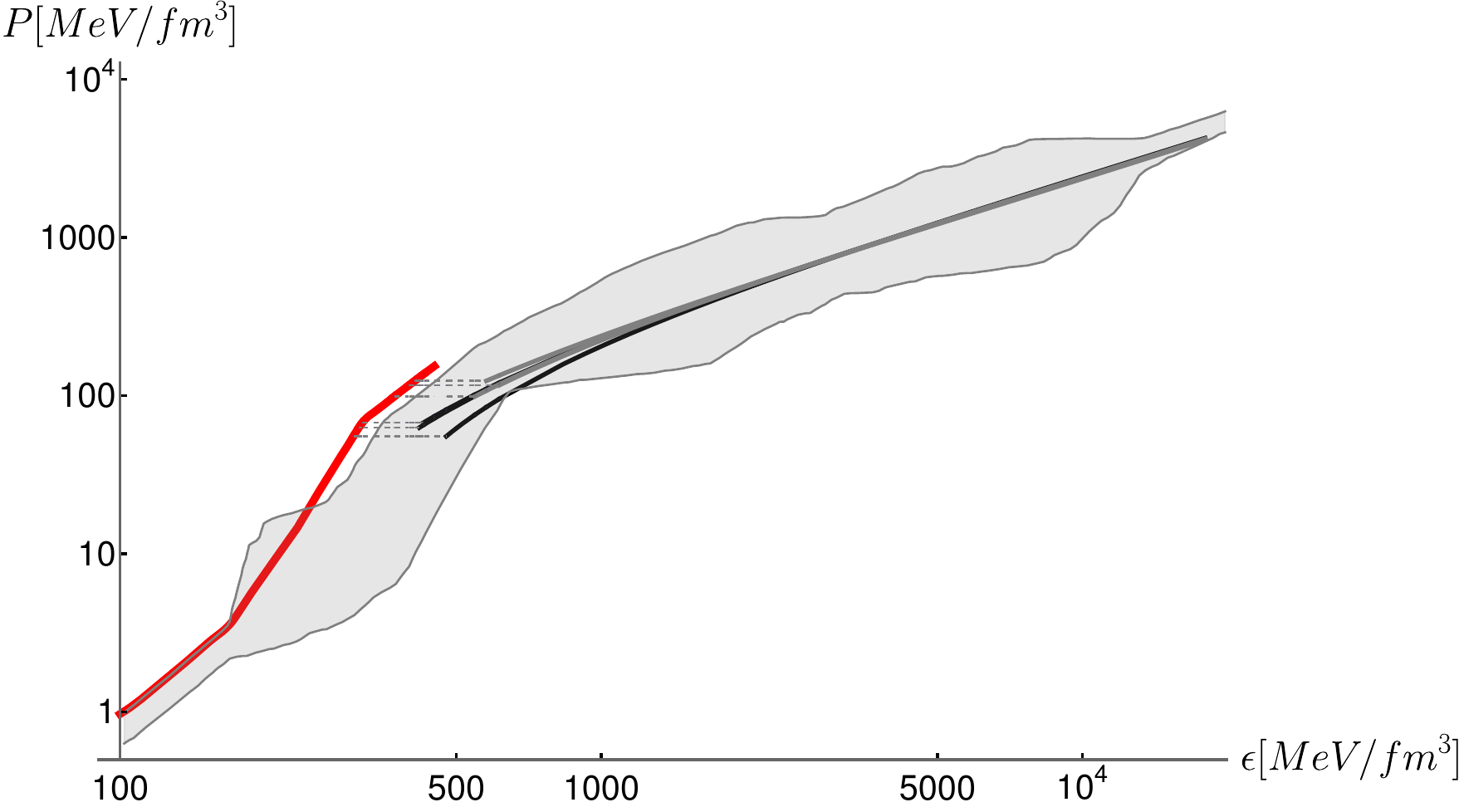} 
    \includegraphics[width=1\linewidth]{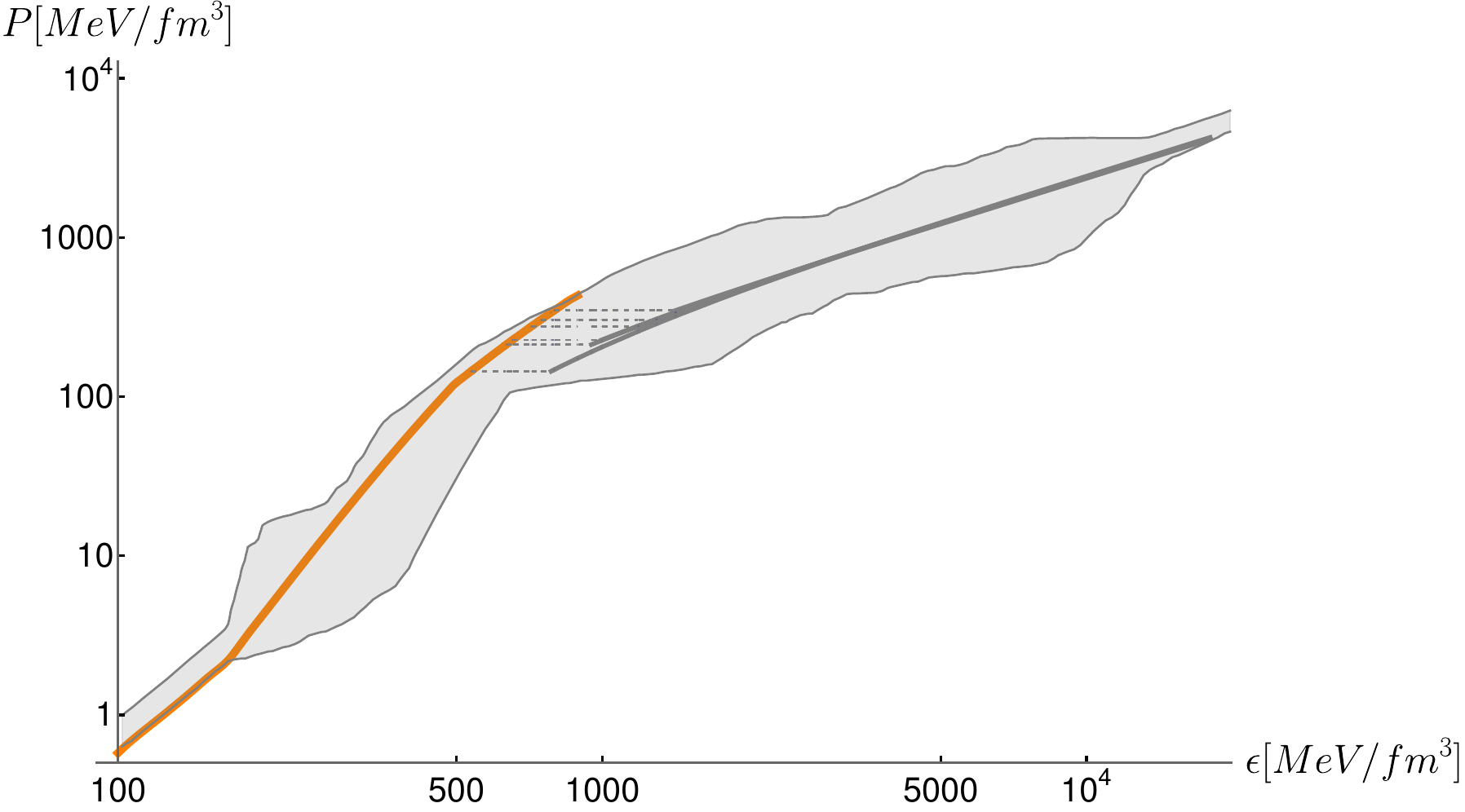}
    \caption{Collection of equations of state in a pressure vs. energy density plot for stiff and medium baryonic phases. Color coding and line styles are as in fig \ref{pressure2}. Phase transitions are indicated by dashed grey lines. The quark phases all asymptote to the pQCD regime. {The gray region found in \cite{Annala:2017llu} is consistent with pQCD and astrophysical constraints.}}
    \label{banana}
\end{figure}
The speed of sound for the 3 parameter values which produce quark stars is shown in figure \ref{fig:speedofsound}. The quark phases are barely distinguishable and approach a value somewhat smaller than the conformal value of $1/3$ in the UV. The speed of sound stays below the speed of light, although the baryonic phase violates the bound from transport $\frac{c_s^2}{c^2}\le 0.781$ found in \cite{Hippert:2024hum}.

\begin{figure}[h]
    \centering
    \includegraphics[width=1\linewidth]{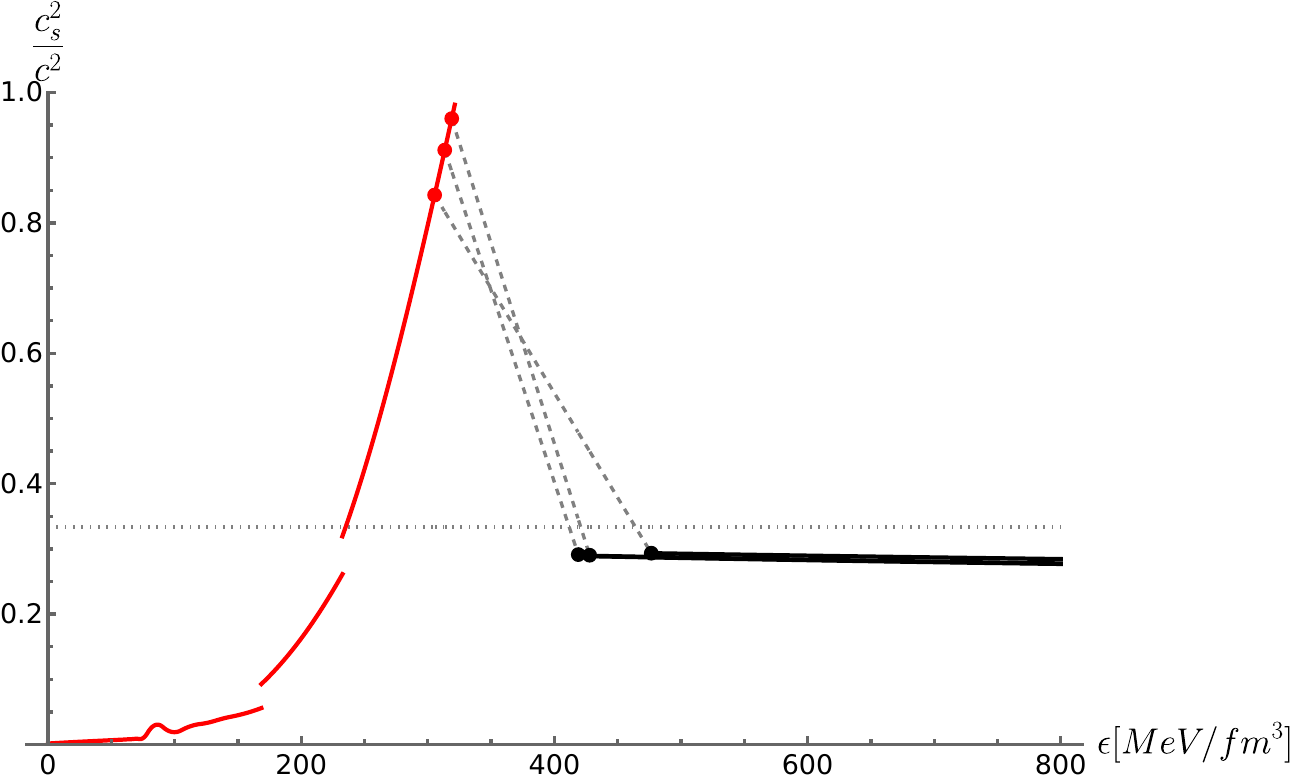} 
    \caption{The speed of sound for the 3 equations of state that allow for stable quark matter cores. Dashed gray lines indicate phase transitions to the quark phases, which in this plot lie on top of each other and approach a value smaller than the conformal $\frac{1}{3}$ (indicated as a vertical dotted gray line) in the UV.}
    \label{fig:speedofsound}
\end{figure}
For those same 3 equations of state we plot the polytropic index 
\begin{align}
    \gamma(\epsilon)=\epsilon \frac{P'(\epsilon)}{P(\epsilon)}
\end{align}
as a function of the normalized density
\begin{align}
    \frac{n(\epsilon)}{n_{\text{sat}}}= \frac{1}{n_{\text{sat}}}\frac{\epsilon+ P(\epsilon)}{\mu_{\text{phys}}(\epsilon)},
\end{align}
where $n_{\text{sat}}$ is the nuclear saturation density, in figure \ref{fig:polytrop}. The polytropic indices in the quark phases clearly exceed $\gamma=1.75$, which was used in \cite{Annala:2019puf, Annala:2021gom} as a criterion for the onset of quark matter. The results shown in figure \ref{fig:polytrop} indicate that this criterion is therefore perhaps too restrictive as in this holographic model quark matter with $\gamma$ as high as $2.5$ can appear. {This is to be contrasted with V-QCD studies, which predict transitions at much larger $\epsilon$ and a smaller polytropic index $\gamma$ in the quark phase \cite{Jokela:2020piw}.}

\begin{figure}[h]
    \centering
    \includegraphics[width=1\linewidth]{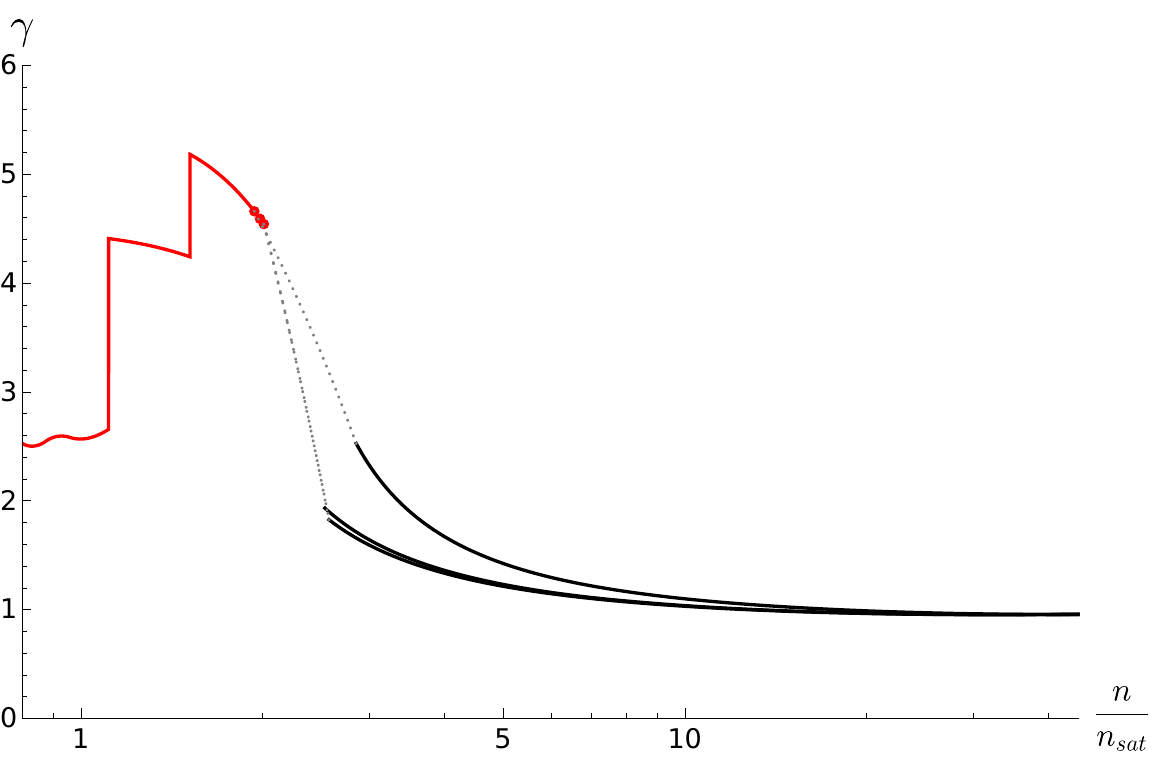} 
    \caption{The polytropic index $\gamma$ for the 3 equations of state that allow for stable quark matter cores as a function of the normalised density $\frac{n}{n_{\text{sat}}}$. Dashed gray lines indicate phase transitions to the quark phases. At large densities, the graphs are slightly below the conformal value of $\gamma=1$.}
    \label{fig:polytrop}
\end{figure}

\subsection{Mass-radius relation}
\label{subsec:massradd}

The mass-radius relation for non-rotating compact stars is determined from the EoS by solving the Tolman-Oppenheimer-Volko{ff} (TOV) equations  \cite{Haensel:2007yy}
\begin{eqnarray}
\frac{dP}{dr}&=&-G_N\left( \epsilon+ P\right)\frac{M+ 4\pi r^3P}{r(r-2G_N M)},\\
\frac{dM}{dr}&=&4\pi r^2\epsilon .\qquad
\end{eqnarray} 
Here $M$ and $P$ are the mass and pressure in the star as a function of radius $r$ and $G_N$ is Newton's constant. To be able to solve the equations we need the EoS $\epsilon(P)$, as well as the central pressure $P_c=P(r = 0)$ as initial condition. The output are the mass $M(r)$ and pressure $P(r)$ of the corresponding star. The radius $R$ of the star will be the value of $r$ at which the pressure vanishes.

The results for both the stiff and medium equations of state supplemented by the holographic quark phase are shown in figure \ref{massvsrad}. Initially (at large radii) the stars are fully composed of nucleons. For any fixed parameter choice, quark cores develop when changing from the red (or orange) baryonic curves to the black (or gray) quark curves. A necessary condition for stability is $\frac{\partial M(\epsilon_c)}{\partial\epsilon_c} >0$, where $\epsilon_c$ is the energy density at the center of the star. The stars, whose quark phase is shown in black, obey this criterion until the maximum in the mass  vs. radius plot. It is thus possible, with the stiff baryonic EoS, to obtain stable quark stars. See also \cite{Albino:2025puc} for a recent analysis of stars with quark cores. When the maximum is passed (from larger to smaller radii), a radial mode becomes unstable and the star will collapse to a black hole. The maximally attainable masses for quark stars in this model for the 3 parameter choices we display are ${M_{\text{max}}}/{M_\odot}=2.17,2.13,1.90$, the first of which is in good agreement (although somewhat favoring lower values) with the currently highest measured mass to date\footnote{PSR J0952-0607 has a rotation period of $1.41$ ms, whilst our analysis assumed no rotation. Rotating compact objects may exceed bounds valid in the non-rotating case, although for rotation periods of the order of ms only a change of order $1-2 \;\%$ in the maximal mass is expected.} of $\frac{M^{\text{exp}}_{\text{max}}}{M_\odot}=2.35 \pm0.17$. 
From highest to lowest maximal mass, the parameters for the stable quark cores are $(R_{\text{AdS}}\;[\text{MeV}^{-1}],\lambda_t)=(0.0175,2.4),(0.0172,2.33),(0.015,1.9)$.
\begin{figure}[h]
    \centering
   
    \includegraphics[width=1\linewidth]{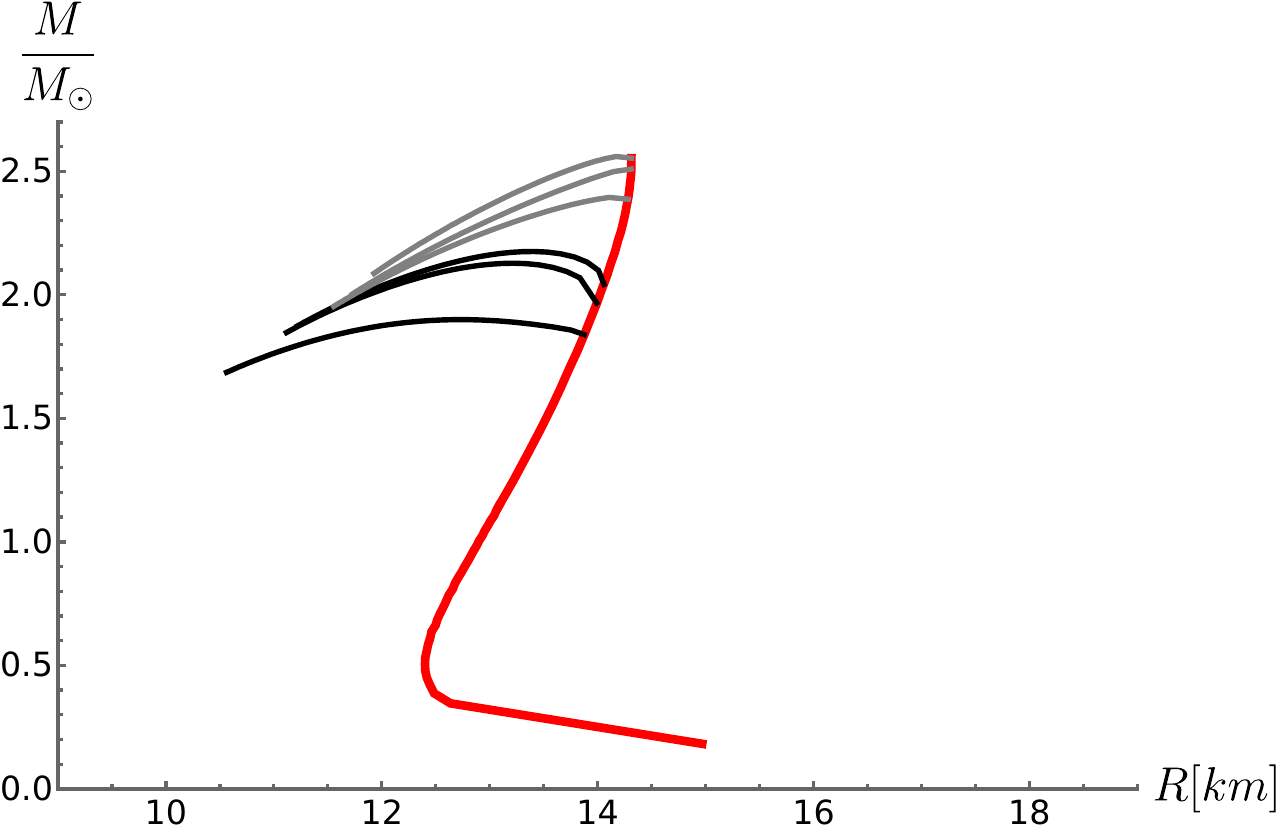}
    \includegraphics[width=1\linewidth]{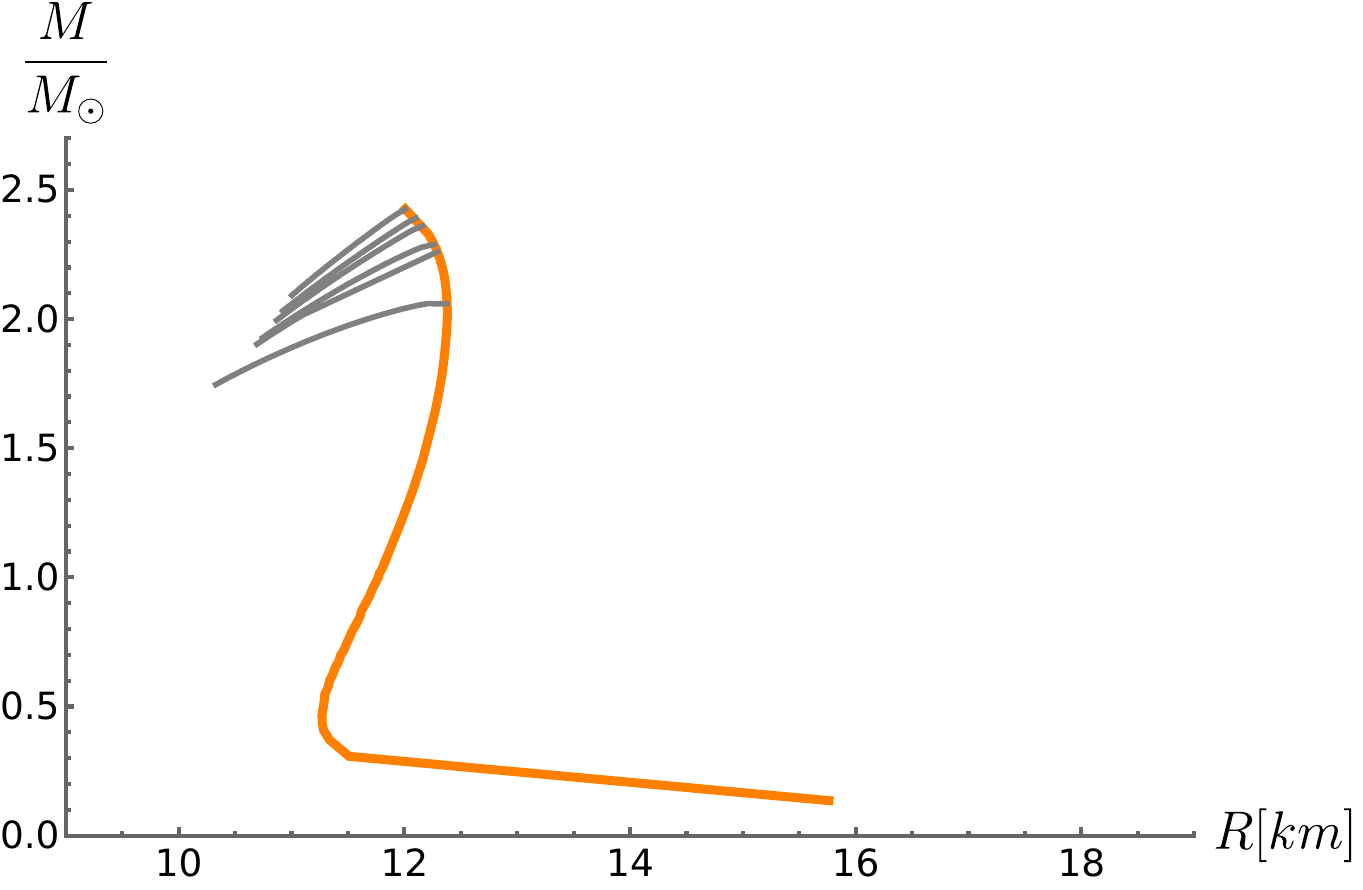}
    \caption{Mass vs radius curves for stiff (top) and medium (bottom) phenomenological baryon phases together with the holographic quark phase. The quark phases highlighted in gray do not lead to stable quark matter in the core, whilst the 3 curves in black in the upper plot support quark stars as high as $M=2.17\, M_\odot$. As mentioned in the text, the stiff baryonic phase is crucial. }
    \label{massvsrad}
\end{figure}

\subsection{Tidal deformability and LIGO constraints}

It is expected that, in a binary system of neutron stars during the inspiral process, the tidal forces between the two stars would have measurable effects in the gravitational wave signal that can be observed with gravitational wave detectors. An observable that parametrizes the effect of tidal forces on a neutron star is the tidal deformability $\bar{\lambda}^{(\mathrm{tid})}$. It connects the EoS that describes the matter inside neutron stars to the gravitational wave emission during the inspiral. It has been shown that a small tidal signature arises in the inspiral below $400$ Hz \cite{Flanagan:2007ix}. This signature amounts to a phase correction which can be described in terms of a single EoS dependent tidal deformability parameter $\bar{\lambda}^{(\mathrm{tid})}$, which is the ratio of each star's induced quadrupole moment {due} to the tidal field of its companion in the binary system.

We follow references \cite{Hinderer:2009ca,Postnikov:2010yn,Zhao:2018nyf} to calculate the tidal deformability. In particular it is important to accurately match correctly between the baryonic part of the star and the quark core as first order phase transitions imply special boundary conditions \cite{Takatsy:2020bnx,Han:2018mtj}.
The dimensionless tidal deformability is defined as
$$\Lambda=\frac{\bar{\lambda}^{(\mathrm{tid})} c^{10}}{G_N^4 M^5}.$$
We plot our results in figure \ref{tidaldeforma}.
In \cite{GuerraChaves:2019foa} $\Lambda$ is obtained for the case of effective models of baryonic and strange quark matter. Qualitatively the confined matter curve (red) in Fig. \ref{tidaldeforma} has a similar behavior as the ones obtained from the models in \cite{GuerraChaves:2019foa}. Once the star gets heavy enough to support quark cores, the tidal deformability decreases rather quickly.
It is however difficult to make quantitative predictions as due to the very stiff baryonic phase, the tidal deformability already considerably overshoots the constraint $\Lambda(1.4 M_\odot)=190 ^{+390}_{-120}$ of \cite{LIGOScientific:2018cki}. The stiff baryonic phase yields $\Lambda(1.4 M_\odot)\sim 950$ and predictions about $\Lambda$ in the quark core regime are affected by this overestimation in the nucleon phase. Upon formation of quark cores until the maximal allowed mass of the star, the tidal deformability drops by $\sim 80$.

\begin{figure}
    \centering
          \includegraphics[width=1 \linewidth]{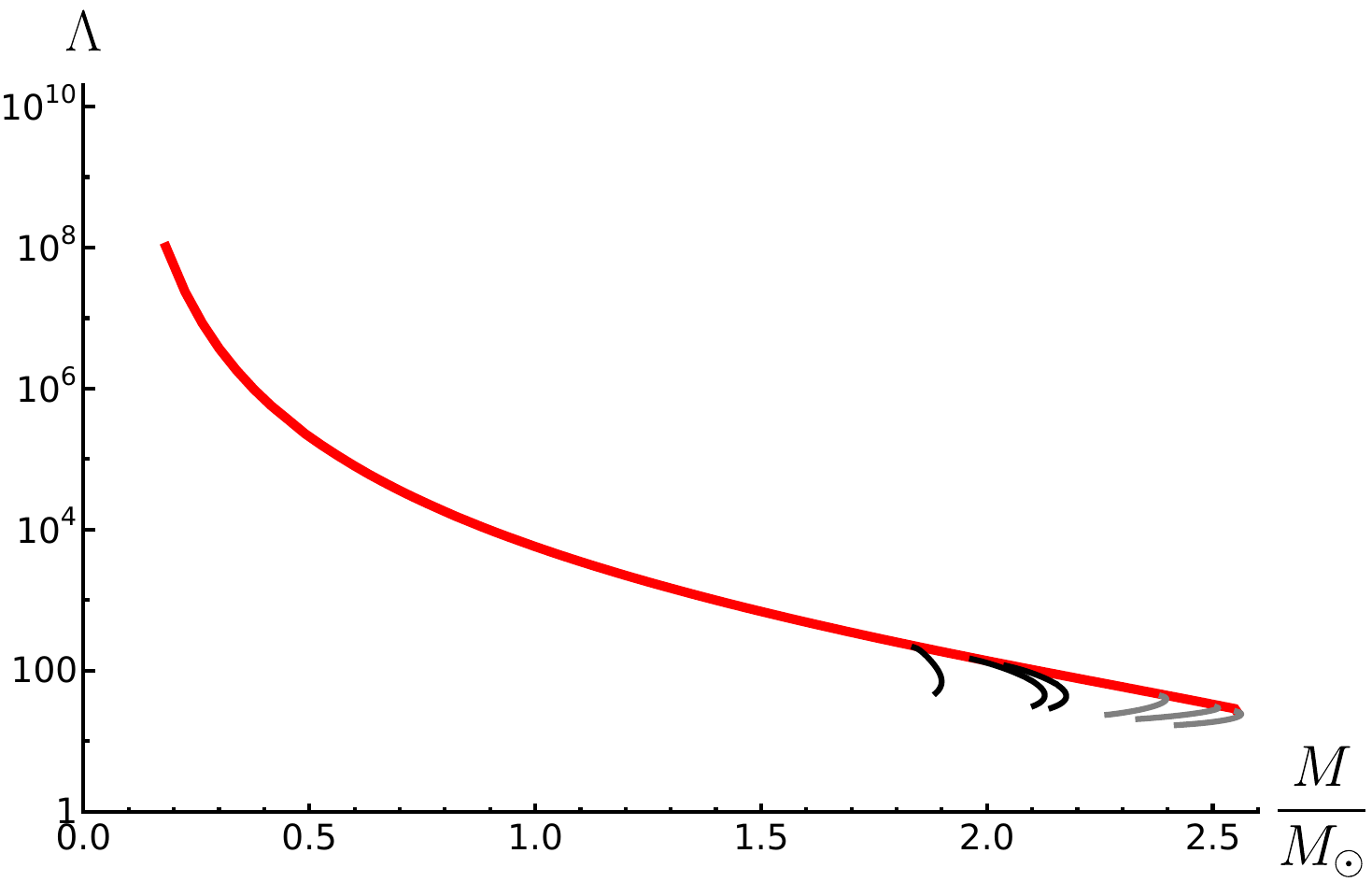}
     \caption{The dimensionless tidal deformability $\Lambda$ as a function of the mass of the star. After the transition from the phenomenological stiff phase (in red) to the quark matter in the core (in black) the tidal deformability decreases rather rapidly.}
    
    \label{tidaldeforma}
\end{figure}

\section{Conclusions}\label{SectionIV}

{In this paper we have investigated the D3/D7 model with a modified dilaton profile. The profiles considered interpolate monotonically and smoothly from 0 in the UV to a finite regular value in the IR. 
Depending on the precise shape of this profile such configurations also allow for baryonic matter in the form of wrapped D5-branes. At least in the smeared approximation the equation of state is however way too stiff and disagrees with phenomenology. Simple modifications of the action or dilaton profile are not able to improve the situation. Due to the absence of a good holographical description of the baryonic phase (at least within this D3/D7 setup) a hybrid approach, utilizing the phenomenological equations of state from \cite{Hebeler:2013nza} and the holographic quark phase, was adopted. These hybrid EoS are largely consistent with existing constraints and for the stiff case are able to produce stable quark stars contrary to most other holographic studies. Furthermore the holographic quark phase produces a polytropic index $\gamma>1.75$ for a region of densities just after the transition. The quark matter criterion of \cite{Annala:2019puf,Annala:2021gom} is thus violated.

The maximal allowed masses are quite close to experimental data, hinting at the possibility that stable quark cores may exist within detected neutron stars. The tidal deformability shows a quick decrease once quark cores appear. Despite some tensions with existing data, specifically with the tidal deformability constraint \cite{LIGOScientific:2018cki}, this simple model allows one to qualitatively study the implications of a first order transition to quark matter within neutron stars. A baryonic EoS in between the stiff and medium phenomenological equations of state fully compatible with existing constraints together with the quark EoS of this paper would likely lead to stable quark cores as well. Interesting observables for future studies include neutrino transport and properties of rotating neutron stars with quark cores. 

The properties of neutron stars largely on the details of the baryonic phase, hence it would be very interesting to study localized D5-brane embeddings within this model, which presumably more accurately describe the low density regime. Moreover only a rather restrictive region of the parameter space for the quark phase was explored and a more extensive analysis is left for future work.

}

\bigskip \noindent {\bf Acknowledgements:} We thank Lorenzo Bartolini and Anton Rebhan for fruitful discussions and comments on the manuscript. J.M wishes to thank Nick Evans and the University of Southampton for their hospitality, where this work was initiated. K.B.F would like to thank the Beijing Institute of Mathematical Sciences and Applications (BIMSA) for their
 hospitality, where part of this work was completed. J.C.R. was supported by a DGAPA-UNAM postdoctoral fellowship and by a SECIHTI postdoctoral fellowship CVU:609042. J.M. has been supported by the Austrian Science Fund FWF, Grant-DOI 10.55776/PAT7221623.

\newpage 
\appendix
\section{Pure D3/D7 quark phase}\label{appendixA}
Let us {briefly} review the quark phase {found in the $\mathcal{N}=4$ duality with $N_f$ D7-branes \cite{Karch:2002sh} and worked out} in \cite{Hoyos:2016zke}. The DBI action for a probe D7-brane in pure AdS with a constant dilaton is
\begin{equation}
S \sim - \int d \rho \rho^3 \sqrt{1 + (\partial_\rho \chi)^2 - 2 \pi \alpha' (\partial_\rho A_t)^2} \end{equation}
The brane embedding function $\chi(\rho)$ is holographically dual to the quark mass and condensate and $A_t$ is a gauge field dual to the quark number chemical potential and density. An analytic form for the grand canonical potential density can be found \cite{Karch:2007br}
\begin{equation}  \label{HoyosF} \Omega \sim {} -(\mu_{\text{phys}}^2 - m^2)^2  + {\cal O}(\mu_{\text{phys}}^3T, T^4)\end{equation}
where $m$ and $\mu_{\text{phys}}$ are the asymptotic values of $\chi$ and $A_t$ respectively. The string length $l_s$ or $\alpha^\prime$ (which is formally zero in the supergravity limit) {actually} cancels from the resulting potential density for the field theory, as usual in the AdS/CFT correspondence.

{By suitably fixing $\lambda_t$ one can} match the asymptotic pQCD {result}
\begin{equation}  \Omega = - {N_c N_f \over 12 \pi^2} \mu_{\text{phys}}^4. \label{uv} \end{equation}

{The parameter $m$ is in principle free and describes quark masses. It has to be bigger than $308$ MeV, otherwise the vacuum phase transitions to a quark phase at $308$ MeV. This parameter space was explored in \cite{Hoyos:2016zke,Annala:2017tqz}, where no stable quark stars were found.}

\bibliographystyle{JHEP}
\bibliography{refs}

@article{Bartolini:2022gdf,
    author = "Bartolini, Lorenzo and Gudnason, Sven Bjarke",
    title = "{Symmetry energy in holographic QCD}",
    eprint = "2209.14309",
    archivePrefix = "arXiv",
    primaryClass = "hep-ph",
    doi = "10.21468/SciPostPhys.16.6.156",
    journal = "SciPost Phys.",
    volume = "16",
    number = "6",
    pages = "156",
    year = "2024"
}

@article{Tootle:2022pvd,
    author = {Tootle, Samuel and Ecker, Christian and Topolski, Konrad and Demircik, Tuna and J{\"a}rvinen, Matti and Rezzolla, Luciano},
    title = "{Quark formation and phenomenology in binary neutron-star mergers using V-QCD}",
    eprint = "2205.05691",
    archivePrefix = "arXiv",
    primaryClass = "astro-ph.HE",
    reportNumber = "APCTP Pre2022 - 007",
    doi = "10.21468/SciPostPhys.13.5.109",
    journal = "SciPost Phys.",
    volume = "13",
    pages = "109",
    year = "2022"
}

@article{Annala:2017llu,
    author = "Annala, Eemeli and Gorda, Tyler and Kurkela, Aleksi and Vuorinen, Aleksi",
    title = "{Gravitational-wave constraints on the neutron-star-matter Equation of State}",
    eprint = "1711.02644",
    archivePrefix = "arXiv",
    primaryClass = "astro-ph.HE",
    reportNumber = "CERN-TH-2017-236",
    doi = "10.1103/PhysRevLett.120.172703",
    journal = "Phys. Rev. Lett.",
    volume = "120",
    number = "17",
    pages = "172703",
    year = "2018"
}

@article{Albino:2025puc,
    author = "Albino, Milena and Malik, Tuhin and Ferreira, M{\'a}rcio and Provid{\^e}ncia, Constan{\c{c}}a",
    title = "{A Bayesian Inference of Hybrid Stars with Large Quark Cores}",
    eprint = "2511.02653",
    archivePrefix = "arXiv",
    primaryClass = "nucl-th",
    month = "11",
    year = "2025"
}

@article{LIGOScientific:2018cki,
    author = "Abbott, B. P. and others",
    collaboration = "LIGO Scientific, Virgo",
    title = "{GW170817: Measurements of neutron star radii and equation of state}",
    eprint = "1805.11581",
    archivePrefix = "arXiv",
    primaryClass = "gr-qc",
    reportNumber = "LIGO-P1800115",
    doi = "10.1103/PhysRevLett.121.161101",
    journal = "Phys. Rev. Lett.",
    volume = "121",
    number = "16",
    pages = "161101",
    year = "2018"
}

@article{Takatsy:2020bnx,
    author = "Tak{\'a}tsy, J{\'a}nos and Kov{\'a}cs, P{\'e}ter",
    title = "{Comment on ''Tidal Love numbers of neutron and self-bound quark stars''}",
    eprint = "2007.01139",
    archivePrefix = "arXiv",
    primaryClass = "astro-ph.HE",
    doi = "10.1103/PhysRevD.102.028501",
    journal = "Phys. Rev. D",
    volume = "102",
    number = "2",
    pages = "028501",
    year = "2020"
}

@article{Leutgeb:2022lqw,
    author = "Leutgeb, Josef and Mager, Jonas and Rebhan, Anton",
    title = "{Hadronic light-by-light contribution to the muon g-2 from holographic QCD with solved U(1)A problem}",
    eprint = "2211.16562",
    archivePrefix = "arXiv",
    primaryClass = "hep-ph",
    doi = "10.1103/PhysRevD.107.054021",
    journal = "Phys. Rev. D",
    volume = "107",
    number = "5",
    pages = "054021",
    year = "2023"
}

@article{Leutgeb:2024rfs,
    author = "Leutgeb, Josef and Mager, Jonas and Rebhan, Anton",
    title = "{Superconnections in AdS/QCD and the hadronic light-by-light contribution to the muon g-2}",
    eprint = "2411.10432",
    archivePrefix = "arXiv",
    primaryClass = "hep-ph",
    doi = "10.1103/PhysRevD.111.114001",
    journal = "Phys. Rev. D",
    volume = "111",
    number = "11",
    pages = "114001",
    year = "2025"
}

@article{Cappiello:2025fyf,
    author = "Cappiello, Luigi and Leutgeb, Josef and Mager, Jonas and Rebhan, Anton",
    title = "{Tensor meson transition form factors in holographic QCD and the muon g {\ensuremath{-}} 2}",
    eprint = "2501.09699",
    archivePrefix = "arXiv",
    primaryClass = "hep-ph",
    doi = "10.1007/JHEP07(2025)033",
    journal = "JHEP",
    volume = "07",
    pages = "033",
    year = "2025"
}

@article{Mager:2025pvz,
    author = "Mager, Jonas and Cappiello, Luigi and Leutgeb, Josef and Rebhan, Anton",
    title = "{Longitudinal Short-Distance Constraints on Hadronic Light-by-Light Scattering and Tensor-Meson Contributions to the Muon g-2}",
    eprint = "2501.19293",
    archivePrefix = "arXiv",
    primaryClass = "hep-ph",
    doi = "10.1103/dxwr-gpsl",
    journal = "Phys. Rev. Lett.",
    volume = "135",
    number = "9",
    pages = "091901",
    year = "2025"
}

@article{Bartolini:2023wis,
    author = "Bartolini, Lorenzo and Gudnason, Sven Bjarke",
    title = "{Neutron stars in the Witten-Sakai-Sugimoto model}",
    eprint = "2307.11886",
    archivePrefix = "arXiv",
    primaryClass = "hep-ph",
    doi = "10.1007/JHEP11(2023)209",
    journal = "JHEP",
    volume = "11",
    pages = "209",
    year = "2023"
}

@article{Han:2018mtj,
    author = "Han, Sophia and Steiner, Andrew W.",
    title = "{Tidal deformability with sharp phase transitions in (binary) neutron stars}",
    eprint = "1810.10967",
    archivePrefix = "arXiv",
    primaryClass = "nucl-th",
    doi = "10.1103/PhysRevD.99.083014",
    journal = "Phys. Rev. D",
    volume = "99",
    number = "8",
    pages = "083014",
    year = "2019"
}

@article{Hippert:2024hum,
    author = "Hippert, Mauricio and Noronha, Jorge and Romatschke, Paul",
    title = "{Upper bound on the speed of sound in nuclear matter from transport}",
    eprint = "2402.14085",
    archivePrefix = "arXiv",
    primaryClass = "nucl-th",
    doi = "10.1016/j.physletb.2024.139184",
    journal = "Phys. Lett. B",
    volume = "860",
    pages = "139184",
    year = "2025"
}

@article{Henningson:1998gx,
    author = "Henningson, M. and Skenderis, K.",
    title = "{The Holographic Weyl anomaly}",
    eprint = "hep-th/9806087",
    archivePrefix = "arXiv",
    reportNumber = "CERN-TH-98-188, KUL-TF-98-21",
    doi = "10.1088/1126-6708/1998/07/023",
    journal = "JHEP",
    volume = "07",
    pages = "023",
    year = "1998"
}

@article{Bartolini:2025sag,
    author = {Bartolini, Lorenzo and Gudnason, Sven Bjarke and J{\"a}rvinen, Matti},
    title = "{Isospin asymmetry and neutron stars in holographic QCD in the Veneziano limit}",
    eprint = "2504.01758",
    archivePrefix = "arXiv",
    primaryClass = "hep-ph",
    reportNumber = "APCTP Pre2025 - 008",
    doi = "10.1103/PhysRevD.111.106021",
    journal = "Phys. Rev. D",
    volume = "111",
    number = "10",
    pages = "106021",
    year = "2025"
}

@article{Jarvinen:2023jbr,
    author = "Jarvinen, Matti",
    title = "{Holographic baryons, dense matter and neutron star mergers}",
    eprint = "2307.01745",
    archivePrefix = "arXiv",
    primaryClass = "hep-ph",
    reportNumber = "APCTP Pre2023 - 007",
    doi = "10.22128/jhap.2023.695.1054",
    journal = "JHAP",
    volume = "3",
    number = "3",
    pages = "1--22",
    year = "2023"
}

@article{Jokela:2020piw,
    author = {Jokela, Niko and J{\"a}rvinen, Matti and Nijs, Govert and Remes, Jere},
    title = "{Unified weak and strong coupling framework for nuclear matter and neutron stars}",
    eprint = "2006.01141",
    archivePrefix = "arXiv",
    primaryClass = "hep-ph",
    reportNumber = "HIP-2020-16/TH",
    doi = "10.1103/PhysRevD.103.086004",
    journal = "Phys. Rev. D",
    volume = "103",
    number = "8",
    pages = "086004",
    year = "2021"
}

@article{Ecker:2019xrw,
    author = {Ecker, Christian and J{\"a}rvinen, Matti and Nijs, Govert and van der Schee, Wilke},
    title = "{Gravitational waves from holographic neutron star mergers}",
    eprint = "1908.03213",
    archivePrefix = "arXiv",
    primaryClass = "astro-ph.HE",
    doi = "10.1103/PhysRevD.101.103006",
    journal = "Phys. Rev. D",
    volume = "101",
    number = "10",
    pages = "103006",
    year = "2020"
}

@article{Chesler:2019osn,
    author = "Chesler, Paul M. and Jokela, Niko and Loeb, Abraham and Vuorinen, Aleksi",
    title = "{Finite-temperature Equations of State for Neutron Star Mergers}",
    eprint = "1906.08440",
    archivePrefix = "arXiv",
    primaryClass = "astro-ph.HE",
    reportNumber = "HIP-2019-17/TH",
    doi = "10.1103/PhysRevD.100.066027",
    journal = "Phys. Rev. D",
    volume = "100",
    number = "6",
    pages = "066027",
    year = "2019"
}

@article{BitaghsirFadafan:2018uzs,
    author = "Bitaghsir Fadafan, Kazem and Kazemian, Farideh and Schmitt, Andreas",
    title = "{Towards a holographic quark-hadron continuity}",
    eprint = "1811.08698",
    archivePrefix = "arXiv",
    primaryClass = "hep-ph",
    doi = "10.1007/JHEP03(2019)183",
    journal = "JHEP",
    volume = "03",
    pages = "183",
    year = "2019"
}

@article{Kovensky:2020xif,
    author = "Kovensky, Nicolas and Schmitt, Andreas",
    title = "{Holographic quarkyonic matter}",
    eprint = "2006.13739",
    archivePrefix = "arXiv",
    primaryClass = "hep-th",
    doi = "10.1007/JHEP09(2020)112",
    journal = "JHEP",
    volume = "09",
    pages = "112",
    year = "2020"
}

@article{Elliot-Ripley:2016uwb,
    author = "Elliot-Ripley, Matthew and Sutcliffe, Paul and Zamaklar, Marija",
    title = "{Phases of kinky holographic nuclear matter}",
    eprint = "1607.04832",
    archivePrefix = "arXiv",
    primaryClass = "hep-th",
    reportNumber = "DCPT-16-23",
    doi = "10.1007/JHEP10(2016)088",
    journal = "JHEP",
    volume = "10",
    pages = "088",
    year = "2016"
}

@article{Kaplunovsky:2012gb,
    author = "Kaplunovsky, Vadim and Melnikov, Dmitry and Sonnenschein, Jacob",
    title = "{Baryonic Popcorn}",
    eprint = "1201.1331",
    archivePrefix = "arXiv",
    primaryClass = "hep-th",
    reportNumber = "UTTG-01-12, ITEP-TH-61-11, TAUP-2940-11",
    doi = "10.1007/JHEP11(2012)047",
    journal = "JHEP",
    volume = "11",
    pages = "047",
    year = "2012"
}

@article{Kim:2007vd,
    author = "Kim, Keun-Young and Sin, Sang-Jin and Zahed, Ismail",
    title = "{Dense holographic QCD in the Wigner-Seitz approximation}",
    eprint = "0712.1582",
    archivePrefix = "arXiv",
    primaryClass = "hep-th",
    doi = "10.1088/1126-6708/2008/09/001",
    journal = "JHEP",
    volume = "09",
    pages = "001",
    year = "2008"
}

@article{Rozali:2007rx,
    author = "Rozali, Moshe and Shieh, Hsien-Hang and Van Raamsdonk, Mark and Wu, Jackson",
    title = "{Cold Nuclear Matter In Holographic QCD}",
    eprint = "0708.1322",
    archivePrefix = "arXiv",
    primaryClass = "hep-th",
    doi = "10.1088/1126-6708/2008/01/053",
    journal = "JHEP",
    volume = "01",
    pages = "053",
    year = "2008"
}

@article{Bergman:2007wp,
    author = "Bergman, Oren and Lifschytz, Gilad and Lippert, Matthew",
    title = "{Holographic Nuclear Physics}",
    eprint = "0708.0326",
    archivePrefix = "arXiv",
    primaryClass = "hep-th",
    doi = "10.1088/1126-6708/2007/11/056",
    journal = "JHEP",
    volume = "11",
    pages = "056",
    year = "2007"
}

@article{Gubser:1999pk,
    author = "Gubser, Steven S.",
    title = "{Dilaton driven confinement}",
    eprint = "hep-th/9902155",
    archivePrefix = "arXiv",
    reportNumber = "HUTP-99-A010",
    month = "2",
    year = "1999"
}

@article{Constable:1999ch,
    author = "Constable, Neil R. and Myers, Robert C.",
    title = "{Exotic scalar states in the AdS / CFT correspondence}",
    eprint = "hep-th/9905081",
    archivePrefix = "arXiv",
    reportNumber = "MCGILL-99-10",
    doi = "10.1088/1126-6708/1999/11/020",
    journal = "JHEP",
    volume = "11",
    pages = "020",
    year = "1999"
}

@article{Chodos:1974je,
    author = "Chodos, A. and Jaffe, R. L. and Johnson, K. and Thorn, Charles B. and Weisskopf, V. F.",
    title = "{A New Extended Model of Hadrons}",
    reportNumber = "MIT-CTP-387-REV, MIT-CTP-387",
    doi = "10.1103/PhysRevD.9.3471",
    journal = "Phys. Rev. D",
    volume = "9",
    pages = "3471--3495",
    year = "1974"
}

@article{ET:2025xjr,
    author = "Abac, Adrian and others",
    collaboration = "ET",
    title = "{The Science of the Einstein Telescope}",
    eprint = "2503.12263",
    archivePrefix = "arXiv",
    primaryClass = "gr-qc",
    reportNumber = "ET-0036C-25",
    month = "3",
    year = "2025"
}

@article{Evans:2021gyd,
    author = "Evans, Matthew and others",
    title = "{A Horizon Study for Cosmic Explorer: Science, Observatories, and Community}",
    eprint = "2109.09882",
    archivePrefix = "arXiv",
    primaryClass = "astro-ph.IM",
    reportNumber = "CE-P2100003-v7, Cosmic Explorer technical report CE-P2100003-v6",
    month = "9",
    year = "2021"
}

@article{Babington:2003vm,
    author = "Babington, J. and Erdmenger, J. and Evans, Nick J. and Guralnik, Z. and Kirsch, I.",
    title = "{Chiral symmetry breaking and pions in nonsupersymmetric gauge / gravity duals}",
    eprint = "hep-th/0306018",
    archivePrefix = "arXiv",
    reportNumber = "HU-EP-03-27, SHEP-03-10",
    doi = "10.1103/PhysRevD.69.066007",
    journal = "Phys. Rev. D",
    volume = "69",
    pages = "066007",
    year = "2004"
}

@article{Ivanenko:1965dg,
    author = "Ivanenko, D. D. and Kurdgelaidze, D. F.",
    title = "{Hypothesis concerning quark stars}",
    doi = "10.1007/BF01042830",
    journal = "Astrophysics",
    volume = "1",
    pages = "251--252",
    year = "1965"
}

@article{Itoh:1970uw,
    author = "Itoh, N.",
    title = "{Hydrostatic Equilibrium of Hypothetical Quark Stars}",
    doi = "10.1143/PTP.44.291",
    journal = "Prog. Theor. Phys.",
    volume = "44",
    pages = "291",
    year = "1970"
}

@article{Alcock:1986hz,
    author = "Alcock, Charles and Farhi, Edward and Olinto, Angela",
    title = "{Strange stars}",
    doi = "10.1086/164679",
    journal = "Astrophys. J.",
    volume = "310",
    pages = "261--272",
    year = "1986"
}

@article{Witten:1984rs,
    author = "Witten, Edward",
    title = "{Cosmic Separation of Phases}",
    reportNumber = "PRINT-84-0400 (IAS,PRINCETON)",
    doi = "10.1103/PhysRevD.30.272",
    journal = "Phys. Rev. D",
    volume = "30",
    pages = "272--285",
    year = "1984"
}

@article{Fraga:2001id,
    author = "Fraga, Eduardo S. and Pisarski, Robert D. and Schaffner-Bielich, Jurgen",
    title = "{Small, dense quark stars from perturbative QCD}",
    eprint = "hep-ph/0101143",
    archivePrefix = "arXiv",
    doi = "10.1103/PhysRevD.63.121702",
    journal = "Phys. Rev. D",
    volume = "63",
    pages = "121702",
    year = "2001"
}

@article{Yagi:2016bkt,
    author = "Yagi, Kent and Yunes, Nicol\'as",
    title = "{Approximate Universal Relations for Neutron Stars and Quark Stars}",
    eprint = "1608.02582",
    archivePrefix = "arXiv",
    primaryClass = "gr-qc",
    doi = "10.1016/j.physrep.2017.03.002",
    journal = "Phys. Rept.",
    volume = "681",
    pages = "1--72",
    year = "2017"
}

@article{Aleixo:2023lue,
    author = "Aleixo, M. and Lenzi, C. H. and de Paula, W. and da Rocha, R.",
    title = "{Quark stars in $D_3$\textendash{}$D_7$ holographic model}",
    eprint = "2310.17719",
    archivePrefix = "arXiv",
    primaryClass = "hep-ph",
    doi = "10.1140/epjc/s10052-024-12619-7",
    journal = "Eur. Phys. J. C",
    volume = "84",
    number = "3",
    pages = "253",
    year = "2024"
}

@article{Hoyos:2016zke,
    author = "Hoyos, Carlos and Rodr\'\i{}guez Fern\'andez, David and Jokela, Niko and Vuorinen, Aleksi",
    title = "{Holographic quark matter and neutron stars}",
    eprint = "1603.02943",
    archivePrefix = "arXiv",
    primaryClass = "hep-ph",
    reportNumber = "FPAUO-16-06, HIP-2016-08-TH",
    doi = "10.1103/PhysRevLett.117.032501",
    journal = "Phys. Rev. Lett.",
    volume = "117",
    number = "3",
    pages = "032501",
    year = "2016"
}

@article{Hebeler:2013nza,
    author = "Hebeler, K. and Lattimer, J. M. and Pethick, C. J. and Schwenk, A.",
    title = "{Equation of state and neutron star properties constrained by nuclear physics and observation}",
    eprint = "1303.4662",
    archivePrefix = "arXiv",
    primaryClass = "astro-ph.SR",
    doi = "10.1088/0004-637X/773/1/11",
    journal = "Astrophys. J.",
    volume = "773",
    pages = "11",
    year = "2013"
}

@article{Kovensky:2021kzl,
    author = "Kovensky, Nicolas and Poole, Aaron and Schmitt, Andreas",
    title = "{Building a realistic neutron star from holography}",
    eprint = "2111.03374",
    archivePrefix = "arXiv",
    primaryClass = "hep-ph",
    doi = "10.1103/PhysRevD.105.034022",
    journal = "Phys. Rev. D",
    volume = "105",
    number = "3",
    pages = "034022",
    year = "2022"
}

@article{Bartolini:2022rkl,
    author = "Bartolini, Lorenzo and Gudnason, Sven Bjarke and Leutgeb, Josef and Rebhan, Anton",
    title = "{Neutron stars and phase diagram in a hard-wall AdS/QCD model}",
    eprint = "2202.12845",
    archivePrefix = "arXiv",
    primaryClass = "hep-th",
    doi = "10.1103/PhysRevD.105.126014",
    journal = "Phys. Rev. D",
    volume = "105",
    number = "12",
    pages = "126014",
    year = "2022"
}

@article{Jokela:2018ers,
    author = {Jokela, Niko and J\"arvinen, Matti and Remes, Jere},
    title = "{Holographic QCD in the Veneziano limit and neutron stars}",
    eprint = "1809.07770",
    archivePrefix = "arXiv",
    primaryClass = "hep-ph",
    reportNumber = "HIP-2018-18/TH",
    doi = "10.1007/JHEP03(2019)041",
    journal = "JHEP",
    volume = "03",
    pages = "041",
    year = "2019"
}

@article{Preis:2016fsp,
    author = "Preis, Florian and Schmitt, Andreas",
    title = "{Layers of deformed instantons in holographic baryonic matter}",
    eprint = "1606.00675",
    archivePrefix = "arXiv",
    primaryClass = "hep-ph",
    doi = "10.1007/JHEP07(2016)001",
    journal = "JHEP",
    volume = "07",
    pages = "001",
    year = "2016"
}

@article{Li:2015uea,
    author = "Li, Si-wen and Schmitt, Andreas and Wang, Qun",
    title = "{From holography towards real-world nuclear matter}",
    eprint = "1505.04886",
    archivePrefix = "arXiv",
    primaryClass = "hep-ph",
    doi = "10.1103/PhysRevD.92.026006",
    journal = "Phys. Rev. D",
    volume = "92",
    number = "2",
    pages = "026006",
    year = "2015"
}

@article{Karch:2007br,
    author = "Karch, Andreas and O'Bannon, Andy",
    title = "{Holographic thermodynamics at finite baryon density: Some exact results}",
    eprint = "0709.0570",
    archivePrefix = "arXiv",
    primaryClass = "hep-th",
    doi = "10.1088/1126-6708/2007/11/074",
    journal = "JHEP",
    volume = "11",
    pages = "074",
    year = "2007"
}

@article{BitaghsirFadafan:2019ofb,
    author = "Bitaghsir Fadafan, Kazem and Cruz Rojas, Jes\'us and Evans, Nick",
    title = "{Deconfined, Massive Quark Phase at High Density and Compact Stars: A Holographic Study}",
    eprint = "1911.12705",
    archivePrefix = "arXiv",
    primaryClass = "hep-ph",
    doi = "10.1103/PhysRevD.101.126005",
    journal = "Phys. Rev. D",
    volume = "101",
    number = "12",
    pages = "126005",
    year = "2020"
}

@book{Haensel:2007yy,
    author = "Haensel, P. and Potekhin, A. Y. and Yakovlev, D. G.",
    title = "{Neutron stars 1: Equation of state and structure}",
    doi = "10.1007/978-0-387-47301-7",
    publisher = "Springer",
    address = "New York, USA",
    volume = "326",
    year = "2007"
}

@article{Gwak:2012ht,
    author = "Gwak, Bogeun and Kim, Minkyoo and Lee, Bum-Hoon and Seo, Yunseok and Sin, Sang-Jin",
    title = "{Holographic D Instanton Liquid and chiral transition}",
    eprint = "1203.4883",
    archivePrefix = "arXiv",
    primaryClass = "hep-th",
    doi = "10.1103/PhysRevD.86.026010",
    journal = "Phys. Rev. D",
    volume = "86",
    pages = "026010",
    year = "2012"
}

@article{Annala:2019puf,
    author = {Annala, Eemeli and Gorda, Tyler and Kurkela, Aleksi and N\"attil\"a, Joonas and Vuorinen, Aleksi},
    title = "{Evidence for quark-matter cores in massive neutron stars}",
    eprint = "1903.09121",
    archivePrefix = "arXiv",
    primaryClass = "astro-ph.HE",
    reportNumber = "CERN-TH-2019-031, HIP-2019-7/TH",
    doi = "10.1038/s41567-020-0914-9",
    journal = "Nature Phys.",
    volume = "16",
    number = "9",
    pages = "907--910",
    year = "2020"
}

@article{Annala:2021gom,
    author = {Annala, Eemeli and Gorda, Tyler and Katerini, Evangelia and Kurkela, Aleksi and N\"attil\"a, Joonas and Paschalidis, Vasileios and Vuorinen, Aleksi},
    title = "{Multimessenger Constraints for Ultradense Matter}",
    eprint = "2105.05132",
    archivePrefix = "arXiv",
    primaryClass = "astro-ph.HE",
    reportNumber = "HIP-2021-11/TH",
    doi = "10.1103/PhysRevX.12.011058",
    journal = "Phys. Rev. X",
    volume = "12",
    number = "1",
    pages = "011058",
    year = "2022"
}

@article{Evans:2010iy,
    author = "Evans, Nick and Gebauer, Astrid and Kim, Keun-Young and Magou, Maria",
    title = "{Holographic Description of the Phase Diagram of a Chiral Symmetry Breaking Gauge Theory}",
    eprint = "1002.1885",
    archivePrefix = "arXiv",
    primaryClass = "hep-th",
    doi = "10.1007/JHEP03(2010)132",
    journal = "JHEP",
    volume = "03",
    pages = "132",
    year = "2010"
}

@article{Jarvinen:2011qe,
    author = "Jarvinen, Matti and Kiritsis, Elias",
    title = "{Holographic Models for QCD in the Veneziano Limit}",
    eprint = "1112.1261",
    archivePrefix = "arXiv",
    primaryClass = "hep-ph",
    reportNumber = "CCTP-2011-30",
    doi = "10.1007/JHEP03(2012)002",
    journal = "JHEP",
    volume = "03",
    pages = "002",
    year = "2012"
}

@article{Annala:2017tqz,
    author = "Annala, Eemeli and Ecker, Christian and Hoyos, Carlos and Jokela, Niko and Rodr\'\i{}guez Fern\'andez, David and Vuorinen, Aleksi",
    title = "{Holographic compact stars meet gravitational wave constraints}",
    eprint = "1711.06244",
    archivePrefix = "arXiv",
    primaryClass = "astro-ph.HE",
    reportNumber = "FPAUO-17/15, HIP-2017-28/TH, FPAUO-17-15, HIP-2017-28-TH",
    doi = "10.1007/JHEP12(2018)078",
    journal = "JHEP",
    volume = "12",
    pages = "078",
    year = "2018"
}

@article{LIGOScientific:2017vwq,
    author = "Abbott, B. P. and others",
    collaboration = "LIGO Scientific, Virgo",
    title = "{GW170817: Observation of Gravitational Waves from a Binary Neutron Star Inspiral}",
    eprint = "1710.05832",
    archivePrefix = "arXiv",
    primaryClass = "gr-qc",
    reportNumber = "LIGO-P170817",
    doi = "10.1103/PhysRevLett.119.161101",
    journal = "Phys. Rev. Lett.",
    volume = "119",
    number = "16",
    pages = "161101",
    year = "2017"
}

@article{Evans:2012cx,
    author = "Evans, Nick and Kim, Keun-Young and Magou, Maria and Seo, Yunseok and Sin, Sang-Jin",
    title = "{The Baryonic Phase in Holographic Descriptions of the QCD Phase Diagram}",
    eprint = "1204.5640",
    archivePrefix = "arXiv",
    primaryClass = "hep-th",
    doi = "10.1007/JHEP09(2012)045",
    journal = "JHEP",
    volume = "09",
    pages = "045",
    year = "2012"
}

@article{Flanagan:2007ix,
    author = "Flanagan, Eanna E. and Hinderer, Tanja",
    title = "{Constraining neutron star tidal Love numbers with gravitational wave detectors}",
    eprint = "0709.1915",
    archivePrefix = "arXiv",
    primaryClass = "astro-ph",
    doi = "10.1103/PhysRevD.77.021502",
    journal = "Phys. Rev. D",
    volume = "77",
    pages = "021502",
    year = "2008"
}

@article{Zhang:2022uin,
    author = "Zhang, Lin and Huang, Mei",
    title = "{Holographic cold dense matter constrained by neutron stars}",
    eprint = "2209.00766",
    archivePrefix = "arXiv",
    primaryClass = "nucl-th",
    doi = "10.1103/PhysRevD.106.096028",
    journal = "Phys. Rev. D",
    volume = "106",
    number = "9",
    pages = "096028",
    year = "2022"
}

@article{Annala:2023cwx,
    author = {Annala, Eemeli and Gorda, Tyler and Hirvonen, Joonas and Komoltsev, Oleg and Kurkela, Aleksi and N{\"a}ttil{\"a}, Joonas and Vuorinen, Aleksi},
    title = "{Strongly interacting matter exhibits deconfined behavior in massive neutron stars}",
    eprint = "2303.11356",
    archivePrefix = "arXiv",
    primaryClass = "astro-ph.HE",
    reportNumber = "HIP-2023-5/TH, HIP-2023-5/TH",
    doi = "10.1038/s41467-023-44051-y",
    journal = "Nature Commun.",
    volume = "14",
    number = "1",
    pages = "8451",
    year = "2023"
}

@article{Demorest:2010bx,
    author = "Demorest, Paul and Pennucci, Tim and Ransom, Scott and Roberts, Mallory and Hessels, Jason",
    title = "{Shapiro Delay Measurement of A Two Solar Mass Neutron Star}",
    eprint = "1010.5788",
    archivePrefix = "arXiv",
    primaryClass = "astro-ph.HE",
    doi = "10.1038/nature09466",
    journal = "Nature",
    volume = "467",
    pages = "1081--1083",
    year = "2010"
}

@article{Antoniadis:2013pzd,
    author = "Antoniadis, John and others",
    title = "{A Massive Pulsar in a Compact Relativistic Binary}",
    eprint = "1304.6875",
    archivePrefix = "arXiv",
    primaryClass = "astro-ph.HE",
    doi = "10.1126/science.1233232",
    journal = "Science",
    volume = "340",
    pages = "6131",
    year = "2013"
}

@article{Fonseca:2016tux,
    author = "Fonseca, Emmanuel and others",
    title = "{The NANOGrav Nine-year Data Set: Mass and Geometric Measurements of Binary Millisecond Pulsars}",
    eprint = "1603.00545",
    archivePrefix = "arXiv",
    primaryClass = "astro-ph.HE",
    doi = "10.3847/0004-637X/832/2/167",
    journal = "Astrophys. J.",
    volume = "832",
    number = "2",
    pages = "167",
    year = "2016"
}

@article{NANOGrav:2019jur,
    author = "Cromartie, H. T. and others",
    collaboration = "NANOGrav",
    title = "{Relativistic Shapiro delay measurements of an extremely massive millisecond pulsar}",
    eprint = "1904.06759",
    archivePrefix = "arXiv",
    primaryClass = "astro-ph.HE",
    doi = "10.1038/s41550-019-0880-2",
    journal = "Nature Astron.",
    volume = "4",
    number = "1",
    pages = "72--76",
    year = "2019"
}

@article{Fonseca:2021wxt,
    author = "Fonseca, E. and others",
    title = "{Refined Mass and Geometric Measurements of the High-mass PSR J0740+6620}",
    eprint = "2104.00880",
    archivePrefix = "arXiv",
    primaryClass = "astro-ph.HE",
    doi = "10.3847/2041-8213/ac03b8",
    journal = "Astrophys. J. Lett.",
    volume = "915",
    number = "1",
    pages = "L12",
    year = "2021"
}

@article{Karch:2002sh,
    author = "Karch, Andreas and Katz, Emanuel",
    title = "{Adding flavor to AdS / CFT}",
    eprint = "hep-th/0205236",
    archivePrefix = "arXiv",
    reportNumber = "UW-PT-02-10",
    doi = "10.1088/1126-6708/2002/06/043",
    journal = "JHEP",
    volume = "06",
    pages = "043",
    year = "2002"
}

@article{Witten:1998zw,
    author = "Witten, Edward",
    editor = "Bergstrom, L. and Lindstrom, U.",
    title = "{Anti-de Sitter space, thermal phase transition, and confinement in gauge theories}",
    eprint = "hep-th/9803131",
    archivePrefix = "arXiv",
    reportNumber = "IASSNS-HEP-98-21",
    doi = "10.4310/ATMP.1998.v2.n3.a3",
    journal = "Adv. Theor. Math. Phys.",
    volume = "2",
    pages = "505--532",
    year = "1998"
}

@article{Sakai:2004cn,
    author = "Sakai, Tadakatsu and Sugimoto, Shigeki",
    title = "{Low energy hadron physics in holographic QCD}",
    eprint = "hep-th/0412141",
    archivePrefix = "arXiv",
    reportNumber = "IU-MSTP-63, YITP-04-70",
    doi = "10.1143/PTP.113.843",
    journal = "Prog. Theor. Phys.",
    volume = "113",
    pages = "843--882",
    year = "2005"
}

@article{Sakai:2005yt,
    author = "Sakai, Tadakatsu and Sugimoto, Shigeki",
    title = "{More on a holographic dual of QCD}",
    eprint = "hep-th/0507073",
    archivePrefix = "arXiv",
    reportNumber = "YITP-05-36, IU-MSTP-71",
    doi = "10.1143/PTP.114.1083",
    journal = "Prog. Theor. Phys.",
    volume = "114",
    pages = "1083--1118",
    year = "2005"
}

@article{Erlich:2005qh,
    author = "Erlich, Joshua and Katz, Emanuel and Son, Dam T. and Stephanov, Mikhail A.",
    title = "{QCD and a holographic model of hadrons}",
    eprint = "hep-ph/0501128",
    archivePrefix = "arXiv",
    reportNumber = "SLAC-PUB-10965, WM-05-101, INT-PUB-05-02",
    doi = "10.1103/PhysRevLett.95.261602",
    journal = "Phys. Rev. Lett.",
    volume = "95",
    pages = "261602",
    year = "2005"
}

@article{DaRold:2005mxj,
    author = "Da Rold, Leandro and Pomarol, Alex",
    title = "{Chiral symmetry breaking from five dimensional spaces}",
    eprint = "hep-ph/0501218",
    archivePrefix = "arXiv",
    reportNumber = "UAB-FT-578",
    doi = "10.1016/j.nuclphysb.2005.05.009",
    journal = "Nucl. Phys. B",
    volume = "721",
    pages = "79--97",
    year = "2005"
}

@article{Karch:2006pv,
    author = "Karch, Andreas and Katz, Emanuel and Son, Dam T. and Stephanov, Mikhail A.",
    title = "{Linear confinement and AdS/QCD}",
    eprint = "hep-ph/0602229",
    archivePrefix = "arXiv",
    reportNumber = "BUHEP-06-02, INT-PUB-06-04",
    doi = "10.1103/PhysRevD.74.015005",
    journal = "Phys. Rev. D",
    volume = "74",
    pages = "015005",
    year = "2006"
}

@article{Nattila:2017wtj,
    author = {N{\"a}ttil{\"a}, J. and Miller, M. C. and Steiner, A. W. and Kajava, J. J. E. and Suleimanov, V. F. and Poutanen, J.},
    title = "{Neutron star mass and radius measurements from atmospheric model fits to X-ray burst cooling tail spectra}",
    eprint = "1709.09120",
    archivePrefix = "arXiv",
    primaryClass = "astro-ph.HE",
    doi = "10.1051/0004-6361/201731082",
    journal = "Astron. Astrophys.",
    volume = "608",
    pages = "A31",
    year = "2017"
}

@article{Miller:2021qha,
    author = "Miller, M. C. and others",
    title = "{The Radius of PSR J0740+6620 from NICER and XMM-Newton Data}",
    eprint = "2105.06979",
    archivePrefix = "arXiv",
    primaryClass = "astro-ph.HE",
    doi = "10.3847/2041-8213/ac089b",
    journal = "Astrophys. J. Lett.",
    volume = "918",
    number = "2",
    pages = "L28",
    year = "2021"
}

@article{Miller:2019cac,
    author = "Miller, M. C. and others",
    title = "{PSR J0030+0451 Mass and Radius from $NICER$ Data and Implications for the Properties of Neutron Star Matter}",
    eprint = "1912.05705",
    archivePrefix = "arXiv",
    primaryClass = "astro-ph.HE",
    doi = "10.3847/2041-8213/ab50c5",
    journal = "Astrophys. J. Lett.",
    volume = "887",
    number = "1",
    pages = "L24",
    year = "2019"
}

@article{BitaghsirFadafan:2020otb,
    author = "Bitaghsir Fadafan, Kazem and Cruz Rojas, Jes{\'u}s and Evans, Nick",
    title = "{Holographic quark matter with colour superconductivity and a stiff equation of state for compact stars}",
    eprint = "2009.14079",
    archivePrefix = "arXiv",
    primaryClass = "hep-ph",
    doi = "10.1103/PhysRevD.103.026012",
    journal = "Phys. Rev. D",
    volume = "103",
    number = "2",
    pages = "026012",
    year = "2021"
}

@article{Kurkela:2010yk,
    author = "Kurkela, Aleksi and Romatschke, Paul and Vuorinen, Aleksi and Wu, Bin",
    title = "{Looking inside neutron stars: Microscopic calculations confront observations}",
    eprint = "1006.4062",
    archivePrefix = "arXiv",
    primaryClass = "astro-ph.HE",
    reportNumber = "BI-TP-2010-17, INT-PUB-10-025",
    month = "6",
    year = "2010"
}

@article{Kopp:2025ggp,
    author = {K{\"o}pp, F{\'a}bio and Horvath, Jorge Ernesto and Hadjimichef, Dimiter and Vasconcellos, C{\'e}sar A. Zen},
    title = "{Quark/Hybrid Stars Within Perturbative QCD in View of the GW170817 Event}",
    doi = "10.1002/asna.20240136",
    journal = "Astron. Nachr.",
    volume = "346",
    number = "3-4",
    pages = "e20240136",
    year = "2025"
}

@article{Jarvinen:2021jbd,
    author = {J{\"a}rvinen, Matti},
    title = "{Holographic modeling of nuclear matter and neutron stars}",
    eprint = "2110.08281",
    archivePrefix = "arXiv",
    primaryClass = "hep-ph",
    reportNumber = "APCTP Pre2021 - 024",
    doi = "10.1140/epjc/s10052-022-10227-x",
    journal = "Eur. Phys. J. C",
    volume = "82",
    number = "4",
    pages = "282",
    year = "2022"
}

@article{Atashi:2024jbk,
    author = "Atashi, Mahdi and Bitaghsir Fadafan, Kazem",
    title = "{Anomalous dimension and quasinormal modes of flavor branes}",
    doi = "10.1016/j.aop.2024.169762",
    journal = "Annals Phys.",
    volume = "469",
    pages = "169762",
    year = "2024"
}

@article{GuerraChaves:2019foa,
    author = "Guerra Chaves, Andreas and Hinderer, Tanja",
    title = "{Probing the equation of state of neutron star matter with gravitational waves from binary inspirals in light of GW170817: a brief review}",
    eprint = "1912.01461",
    archivePrefix = "arXiv",
    primaryClass = "nucl-th",
    doi = "10.1088/1361-6471/ab45be",
    journal = "J. Phys. G",
    volume = "46",
    number = "12",
    pages = "123002",
    year = "2019"
}

@article{Zhao:2018nyf,
    author = "Zhao, Tianqi and Lattimer, James M.",
    title = "{Tidal Deformabilities and Neutron Star Mergers}",
    eprint = "1808.02858",
    archivePrefix = "arXiv",
    primaryClass = "astro-ph.HE",
    doi = "10.1103/PhysRevD.98.063020",
    journal = "Phys. Rev. D",
    volume = "98",
    number = "6",
    pages = "063020",
    year = "2018"
}

@article{Postnikov:2010yn,
    author = "Postnikov, Sergey and Prakash, Madappa and Lattimer, James M.",
    title = "{Tidal Love Numbers of Neutron and Self-Bound Quark Stars}",
    eprint = "1004.5098",
    archivePrefix = "arXiv",
    primaryClass = "astro-ph.SR",
    doi = "10.1103/PhysRevD.82.024016",
    journal = "Phys. Rev. D",
    volume = "82",
    pages = "024016",
    year = "2010"
}

@article{Hinderer:2009ca,
    author = "Hinderer, Tanja and Lackey, Benjamin D. and Lang, Ryan N. and Read, Jocelyn S.",
    title = "{Tidal deformability of neutron stars with realistic equations of state and their gravitational wave signatures in binary inspiral}",
    eprint = "0911.3535",
    archivePrefix = "arXiv",
    primaryClass = "astro-ph.HE",
    doi = "10.1103/PhysRevD.81.123016",
    journal = "Phys. Rev. D",
    volume = "81",
    pages = "123016",
    year = "2010"
}

@article{BitaghsirFadafan:2024icz,
    author = "Bitaghsir Fadafan, Kazem and Gholamzadeh, Mansoureh",
    title = "{Chiral Symmetry Breaking and the Critical Point in QCD-like Theories}",
    doi = "10.22128/jhap.2024.843.1087",
    journal = "JHAP",
    volume = "4",
    number = "3",
    pages = "19--34",
    year = "2024"
}

@article{Ahmadvand:2021wtt,
    author = "Ahmadvand, M. and Ashoorioon, A. and Bitaghsir Fadafan, K.",
    title = "{Confronting the magnetically-induced holographic composite inflation with observation}",
    eprint = "2103.08362",
    archivePrefix = "arXiv",
    primaryClass = "hep-th",
    reportNumber = "IPM/P-2021/007",
    doi = "10.1103/PhysRevD.104.063509",
    journal = "Phys. Rev. D",
    volume = "104",
    number = "6",
    pages = "063509",
    year = "2021"
}

@article{Leutgeb:2025jmv,
    author = "Leutgeb, Josef and Mager, Jonas and Rebhan, Anton",
    title = "{Divergences in the hadronic light-by-light amplitude of the holographic soft-wall model}",
    eprint = "2511.11797",
    archivePrefix = "arXiv",
    primaryClass = "hep-ph",
    month = "11",
    year = "2025"
}

\end{document}